\documentclass[aps,showpacs,amssymb,amsfonts,twocolumn,superscriptadress]{revtex4}
\usepackage{amsmath,bbm,mathrsfs}
\usepackage{graphicx}
\usepackage{amsfonts}
\usepackage{amssymb}
\usepackage{psfrag}

\def\LRA{\mathop{-\!\!\!-\!\!\!\longrightarrow}\nolimits}
\def\LRAA{\mathop{\!\longrightarrow}\nolimits}

\def\textbf#1{{\bf #1}}
\def\be{\begin{equation}}
\def\ee{\end{equation}}
\def\ben{\begin{eqnarray}}
\def\een{\end{eqnarray}}
\def\eea{\end{array}}
\def\bea{\begin{array}}
\newcommand{\ot}[0]{\otimes}
\newcommand{\Tr}[1]{\mathrm{Tr}#1}
\newcommand{\bei}{\begin{itemize}}
\newcommand{\eei}{\end{itemize}}
\newcommand{\ket}[1]{|#1\rangle}
\newcommand{\bra}[1]{\langle#1|}

\newcommand{\proj}[1]{\ket{#1}\!\bra{#1}}

\newcommand{\Ke}[1]{\big|#1\big\rangle}
\newcommand{\Br}[1]{\big< #1\big|}
\newcommand{\mo}[1]{\left|#1\right|}
\newcommand{\nms}{\negmedspace}
\newcommand{\nmss}{\negmedspace\negmedspace}
\newcommand{\nmsss}{\negmedspace\negmedspace\negmedspace}
\newcommand{\norsl}[1]{\left\|#1\right\|_{\mathrm{1}}}

\newcommand{\pr}[1]{\ket{#1}\bra{#1}}
\newcommand{\ke}[1]{|#1\rangle}
\newcommand{\br}[1]{\langle #1|}
\newcommand{\kett}[1]{\ket{#1}}
\newcommand{\braa}[1]{\bra{#1}}

\def\blacksquare{\vrule height 4pt width 3pt depth2pt}

\begin{document}

\title{Multipartite secret key distillation and bound entanglement }

\author{Remigiusz Augusiak}
\email{remigiusz.augusiak@icfo.es} \affiliation{Faculty of Applied
Physics and Mathematics, Gda\'nsk University of Technology,
Narutowicza 11/12, 80--952 Gda\'nsk, Poland}
\affiliation{ICFO--Institute Ci\'encies Fot\'oniques,
Mediterranean Technology Park, 08860 Castelldefels (Barcelona),
Spain}

\author{Pawe{\l} Horodecki}
\email{pawel@mif.pg.gda.pl} \affiliation{Faculty of Applied
Physics and Mathematics, Gda\'nsk University of Technology,
Narutowicza 11/12, 80--952 Gda\'nsk, Poland}

%
\begin{abstract}
Recently it has been shown that quantum cryptography beyond pure
entanglement distillation is possible and a paradigm for the
associated protocols has been established. Here we systematically
generalize the whole paradigm to the multipartite scenario. We provide
constructions of new classes of multipartite bound entangled
states, i.e., those with underlying  twisted GHZ structure and
nonzero distillable cryptographic key. We quantitatively estimate
the key from below with help of the privacy squeezing technique.
\end{abstract}
\maketitle

\section{Introduction}
Quantum cryptography is one of the most successful applications of
quantum physics in information theory. The original pioneering
BB84 scheme \cite{BB84} was based on sending nonorthogonal states
through an insecure quantum channel. Then the alternative approach
(E91) \cite{Ekert} based on generating key from pure entangled
quantum state have been proposed and later extended to the case of
mixed states in quantum privacy amplification scheme \cite{QPA}
which exploited the idea of distillation of pure entangled quantum
states from more copies of noisy entangled (mixed) states
\cite{distillation}. Much later it was realized that actually
existence of (may be noisy) initial entanglement in the state is
necessary for any type of protocols distilling secret key from
quantum states \cite{Curty1,Curty2}. In a meantime the problem of
unconditional security (security in the most unfriendly scenario
when the eavesdropper may apply arbitrarily correlated
measurements on the sent particles or, in the entanglement
distillation scheme, distribute many particles in a single
entangled quantum state) was further solved in Ref. \cite{LoChau}
in terms of entanglement distillation showing equivalence between
the two (BB84 and E91) ideas (see Ref. \cite{ShorPreskill} for
an alternative proof). However, still the protocol worked only for
entanglement that could be distilled. Also, other
protocols \cite{DW1,DW2} that exploited a modern approach to secrecy
(based on classical notions) also were used in cases when pure
entanglement was distillable. It was known, however, for a
relatively long time that there are states (called bound
entangled) that can not be distilled to pure form \cite{bound}. In
the above context it was quite natural to expect that bound
entangled states cannot lead to private key. However, it happens
not to be true \cite{KH0}: one can extend the entanglement
distillation idea from distillation of pure states to distillation
of {\it private states} (in general mixed states that contain
a private bit) and further show that there are examples of bound
entangled states from which secure key can be distilled. A general
paradigm has been systematically worked out in Refs.
\cite{KH,KHPhD} with further examples of bound
entangled states with secure key \cite{PHRANATO,HPHH} and interesting applications
\cite{RenesSmith,RB,Christandl}. From the quantum channels
perspective the extended scheme \cite{KH0} represents secure key
distillation with help of a quantum channel with vanishing quantum
capacity (i.e., it is impossible to transmit qubit states faithfully).
Those channels \cite{KH0,HPHH} were later used in the discovery
\cite{SmithYard} of the drastically non intuitive, fully nonclassical
effect of mutual activation of zero capacity channels, which
"unlock" each other allowing to transmit quantum information
faithfully if encoded into entanglement across two channels
inputs. On the other hand with help of the seminal machinery
exploiting the notion of almost productness in unconditionally
secure quantum key distillation \cite{Renner} it has been shown
that unconditional security under channels that do not convey quantum
information is possible \cite{UnconditionalSecurity}. Here we
would like to stress that we focus on the approach to quantum
cryptography based on private states rather than the, to some extent,
complementary information--theoretic approach which has also been proven
very fruitful (see Refs. \cite{DW1,RK,KGR,MCL,Renner}).

The results discussed above concern bipartite states. The aim of
the present paper, which, among others, concludes part of the
analysis of \cite{DoktoratRA}, is to develop the general approach
to distillation of secure key from multipartite states. Basically
the content of the paper can be divided into two parts. In the first
part we systematically and in a consistent way generalize the approach
from Ref. \cite{KH}. Here the basic notion of multipartite p--dits
that has been introduced and analyzed already in the previous
paper \cite{PHRA}.

It should be stressed here that, as extensively discussed in
\cite{KHPhD}, other modifications of the paradigm are possible as
far as the so--called notion of "direct accessibility of cryptographic
key" is considered. The p--dit approach is based on local von
Neumann measurements, while it is possible also to consider local
POVM--s \cite{RenesSmith}. Both approaches (and additional one)
were proved to be equivalent in terms of the amount of distillable key
contained in a given bipartite state in Ref. \cite{KHPhD}. While
we leave this issue for further analysis we strongly believe that
the abstract proofs of the latter work naturally extend to our
multipartite case.

The first part of the present paper contains qualitatively new
elements like conditions for closeness to a p--dit state which were
not known so far, and a derivation of a lower bound for multipartite
key where an additional analysis of distance to so--called cq states
was needed. The second part of the paper contains constructions of
novel states ie. multipartite states that contain secure key
though are bound entangled. The states are based on the underlying
(twisted) $N$--partite GHZ structure and are PPT under any $N-1$
versus one system partial transpose. The secret key content is
bounded from below quantitatively with help of the technique adopted
form \cite{HPHH}.

More specifically after basic definitions and a generalization of
the modern definition (that has already become standard) of
secure key distillation from quantum state in Sec. II  we pass to Sec. III
where the notion of multipartite private--dit state (in short
p--dit) and its properties are discussed including especially the
condition for $\epsilon-$closeness to a p--dit. Distillable
cryptographic key in terms of p-dits is analyzed in section IV.
Here an upper bound in terms of relative entropy is proved in analogy
to the bipartite case. A lower bound on the key based on
a modification of the one--way Devetak--Winter protocol \cite{DW1,DW2}
to the multipartite case is provided with help of a natural lemma with
a somewhat involved proof. Also the application \cite{HPHH} of
privacy squeezing \cite{KH0} is naturally extended and applied
here.

The next section is the longest one since it contains all the
constructions of multipartite bound entanglement with
cryptographic key. Note that the first construction, being
an extension and modification of bipartite examples from Ref.
\cite{PHRANATO}, requires nontrivial coincidence of several
conditions that are contained in Lemma V.3. They ensure that, on
the one hand the state is PPT, but on the other it allows to be
modified by the LOCC recurrence protocol to a state that is close to
a multipartite p--dit. This is equivalent to distillability.
Independently, a quantitative analysis is performed illustrating
how the lower bound for distillable key becomes positive. The
second class of bound entangled states (to some extent inspired by
bipartite four--qubit states from \cite{HPHH}) involves hermitian
unitary block elements of the density matrix. Here the
construction is different and, in comparison to the first one, the
observed secure key is much stronger. Finally we shortly recall
the limitations of quantum cryptography \cite{RAPH1,RAPH2}.
Section VI contains conclusions.

\section{Basic notions and the standard definition of secure key}

In what follows we shall be concerned with the scenario in
which $N$ parties $A_{1},\ldots,A_{N}$ wish to obtain
perfectly correlated strings of bits (or in general dits)
that are completely uncorrelated to the eavesdropper Eve by means of
local operations and public communication (LOPC). Let us recall that
the difference between the standard local operations and classical communication (LOCC)
and LOPC lies in the fact that in the latter we need to remember that
any classical message announced by the involved parties may be registered
by Eve. Therefore in comparison to the LOCC paradigm in the LOPC
paradigm one also includes the map (see e.g. Refs. \cite{Christandl,KH})
\begin{eqnarray}
\varrho_{AA'BE}&\nmss=\nmss&\sum_{i}\varrho_{ABE}^{(i)}\ot\proj{i}_{A'}\LRAA
\varrho_{AA'BB'EE'}\nonumber\\
&\nmss=\nmss&\sum_{i}\varrho_{ABE}^{(i)}\ot\proj{i}_{A'}
\ot\proj{i}_{B'}\ot\proj{i}_{E'},
\end{eqnarray}
From the quantum cryptographic point of view the common
aim of all the parties $A_{1},\ldots,A_{N}$ is to distill the following state
\begin{equation}\label{idealcq}
\varrho_{\mathsf{A}E}^{(N,\mathrm{id})}=
\frac{1}{d}\sum_{i=0}^{d-1}\proj{e_{i}^{(1)}\ldots e_{i}^{(N)}}\ot
\varrho^{E},
\end{equation}
called hereafter {\it ideal c...cq (\textsf{c}q) state},
by means of LOPC. Here $\mathsf{A}\equiv A_{1}\ldots A_{N}$ and
$\{\ket{e_{i}^{(j)}}\}_{i=0}^{d-1}$ is some orthonormal
basis in the Hilbert space corresponding to the $j$th party
(denoted hereafter by $\mathcal{H}_{j}$). Their tensor product constitutes
the product basis in $\mathcal{H}_{1}\ot\ldots\ot\mathcal{H}_{N}$, which
we shall denote as
\begin{equation}
\mathcal{B}_{N}^{\mathrm{prod}}=
\left\{\ket{e_{i_{1}}^{(1)}}\ot\ldots\ot\ket{e_{i_{N}}^{(N)}}\right\}_{i_{1},\ldots,i_{N}=0}^{d-1}.
\end{equation}
(In what follows we will be often assuming
$\{\ket{e_{i}^{(j)}}\}_{i=0}^{d-1}$ to be the standard basis in
$\mathcal{H}_{j}$.) One sees that the ideal \textsf{c}q states
represent perfect classical correlations with respect to
the product basis $\mathcal{B}_{N}^{\mathrm{prod}}$ that
uncorrelated to the eavesdropper's degrees of freedom.

We may also define a general \textsf{c}q state to be
\begin{equation}\label{nccq}
\varrho_{\mathsf{A}E}^{(N,\textsf{c}q)}=\sum_{i_{1},\ldots,i_{N}=0}^{d-1}
p_{i_{1}\ldots i_{N}}\proj{e^{(1)}_{i_{1}}\ldots e_{i_{N}}^{(N)}}
\ot\varrho_{i_{1}\ldots i_{N}}^{E}.
\end{equation}
In the above considerations formula we could take different dimensions
on each side, however, for simplicity we restrict to the
case of equal dimensions. All the parties should have
strings of the same length at the end of the protocol to make a key.

It should be also emphasized that in what follows the $j$th party
is assumed to have an additional 'garbage' quantum system defined
on some Hilbert space $\mathcal{H}_{j}'$. Thus we will be assuming
that usually the states shared by the parties are defined on the
Hilbert space $\mathcal{H}\ot\mathcal{H}'$, where
$\mathcal{H}=\mathcal{H}_{1}\ot\ldots\ot\mathcal{H}_{N}$ and
$\mathcal{H}'=\mathcal{H}_{1}'\ot\ldots\ot\mathcal{H}_{N}'$,
$\mathcal{B}_{N}^{\mathrm{prod}}$ constitutes the product basis in
$\mathcal{H}$. Also, following Ref. \cite{KH}, the part of a given
state corresponding to $\mathcal{H}$ ($\mathcal{H}'$) will be sometimes called
{\it the key part} ({\it the shield part}). This terminology comes
from the fact that the key part is the one from which the parties
obtain the cryptographic key, while the shield part protects
secret correlation from the eavesdropper.

Following e.g. Refs. \cite{KH,Christandl}, using the notion of
\textsf{c}q states, we may define the distillable cryptographic
key in the multipartite scenario as follows.

{\it Definition II.1.} Let $\varrho_{\mathsf{A}E}$ be a state
acting on
$\mathbb{C}^{d_{1}}\ot\ldots\ot\mathbb{C}^{d_{N}}\ot\mathbb{C}^{d_{E}}$
and $(P_{n})_{n=1}^{\infty}$ be a sequence of LOPC operations such
that $P_{n}(\varrho_{\mathsf{A}E}^{\ot
n})=\varrho_{\mathsf{A}E}^{(\mathsf{c}\mathrm{q},n)}$, where
$\varrho_{\mathsf{A}E}^{(\mathsf{c}\mathrm{q},n)}$ is a
\textsf{c}q state with $\mathsf{A}$ part defined on
$\big(\mathbb{C}^{d_{n}}\big)^{\ot N}$. The set of operations
$P=(\Lambda_{n})_{n=1}^{\infty}$ is said to be a cryptographic
key distillation protocol if
\begin{equation}
\lim_{n\to\infty}\norsl{\varrho_{\mathsf{A}E}^{(\mathsf{c}\mathrm{q},n)}-\varrho_{\mathsf{A}E}^{(\mathrm{id},n)}}=0,
\end{equation}
where $\varrho_{\mathsf{A}E}^{(\mathrm{id},n)}$ is the ideal $\mathsf{c}$q state defined on the same Hilbert space
as $\varrho_{\mathsf{A}E}^{(\mathsf{c}\mathrm{q},n)}$. We define the rate of
the protocol $P=(P_{n})_{n=1}^{\infty}$ as
\begin{equation}
R_{P}(\varrho_{\mathsf{A}E})=\limsup_{n\to\infty}\frac{\log d_{n}}{n}
\end{equation}
and the distillable classical key as
\begin{equation}
C_{D}(\varrho_{\mathsf{A}E})=\sup_{P}R_{P}(\varrho_{\mathsf{A}E}).
\end{equation}
If instead of $\varrho_{\mathsf{A}E}$ one has the purification $\ket{\psi_{\mathsf{A}E}}$
we write $C_{D}(\varrho_{\mathsf{A}})$.

Let us also mention that a good indicator of the secrecy of our correlations
as well as the uniformity of the probability distribution $p_{i_{1}\ldots i_{N}}$
is the trace norm distance $\big\|\varrho_{\mathsf{A}E}^{(\mathrm{id})}-\varrho_{\mathsf{A}E}^{(\textsf{c}q)}\big\|_{1}$.

\section{Private states}
\subsection{Definition and properties}
Here we discuss the multipartite generalizations of two important
concepts of the scheme from Refs. \cite{KH0,KH}.
Firstly we introduce the notion of twisting and then the notion of
multipartite private states.

{\it Definition III.1.} Let $(U_{i_{1}\ldots i_{N}})_{i_{1}\ldots i_{N}}$ be some family of
unitary operations acting on
$\mathcal{H}'$. Given the $N$--partite product basis $\mathcal{B}_{N}^{\mathrm{prod}}$
we define {\it multipartite twisting} to be the unitary operation given by the following
formula
\begin{equation}
U_{t}=\sum_{i_{1},\ldots,i_{N}=0}^{d-1}\pr{e_{i_{1}}^{(1)}\ldots
e_{i_{N}}^{(N)}}\ot U_{i_{1}\ldots i_{N}}.
\end{equation}
This is an important notion since, as shown in the bipartite case in Ref. \cite{KH} (Theorem 1)
and as it holds also for multipartite states, application of twisting
(taken with respect to the product basis $\mathcal{B}_{N}^{\mathrm{prod}}$)
to a given state $\varrho_{\mathsf{AA}'}$ does not have any effect on the
\textsf{c}q state obtained upon a measurement of the $\mathsf{A}$ part of
the purification of $\varrho_{\mathsf{AA}'}$
in the 
product basis $\mathcal{B}_{N}^{\mathrm{prod}}$. More precisely
states $\varrho_{\mathsf{AA}'}$ and $U_{t}\varrho_{\mathsf{AA}'}U_{t}^{\dagger}$
have the same \textsf{c}q state with respect to $\mathcal{B}_{N}^{\mathrm{prod}}$
for any twisting that is constructed using $\mathcal{B}_{N}^{\mathrm{prod}}$.

We can now pass to the notion of multipartite private states.
These are straightforward generalization of private states
from Refs. \cite{KH0,KH} and were defined already in Ref. \cite{PHRA}.

{\it Definition III.2.} Let $U_{i}$ be some unitary operations for every $i$ and
let $\varrho_{\mathsf{A}'}$ be a density matrix acting on $\mathcal{H}'$. By {\it
multipartite private state} or {\it multipartite pdit} we mean the
following
\begin{equation}\label{mpbit}
\Gamma_{\mathsf{AA'}}^{(d)}=\frac{1}{d}\sum_{i,j=0}^{d-1}\ke{e_{i}^{(1)}\ldots
e_{i}^{(N)}}\br{e_{j}^{(1)}\ldots e_{j}^{(N)}}\ot
U_{i}\varrho_{\mathsf{A}'}U_{j}^{\dag}.
\end{equation}
Naturally, for $N=2$ the above reproduces the bipartite private
states $\gamma_{A_{1}A_{2}A_{1}'A_{2}'}^{(d)}$ introduced in Ref.
\cite{KH}. It follows from the definition that any multipartite
private state may be written as
$\Gamma_{\mathsf{AA'}}^{(d)}=U_{t}(P_{d,N}^{(+)}\ot\varrho_{\mathsf{A}'})U_{t}^{\dagger}$
with $\varrho_{\mathsf{A}'}$ and $U_{t}$ denoting some density
matrix acting on $\mathcal{H}'$ and some twisting, respectively.
Moreover, $P_{d,N}^{(+)}$ stands for the projector onto the
so--called $GHZ$ state \cite{GHZstates} given by
\begin{equation}\label{GHZstates}
\ket{\psi_{d,N}^{(+)}}=\sum_{i=0}^{d-1}\ket{i}^{\ot N}.
\end{equation}
In other words we say that multipartite private states are twisted
$GHZ$ states tensored with an arbitrary density matrix
$\varrho_{\mathsf{A}'}$.

As a simple but illustrative example of a multipartite pdit one
may consider the following $(2D)^{N}\times (2D)^{N}$ state (with
$\mathcal{H}=(\mathbb{C}^{2})^{\ot N}$ and
$\mathcal{H}'=(\mathbb{C}^{D})^{\ot N}$)
\begin{eqnarray}
\Gamma_{\mathrm{ex}}^{(2)}&=&\frac{1}{2D^{N}} \left[
\begin{array}{cccc}
\mathbbm{1}_{D^{N}} & 0 & \ldots &
V_{\pi}^{(D)}\\
0 & 0 & \ldots  & 0\\
\vdots & \vdots & \ddots & \vdots \\
V_{\pi}^{(D)\dagger} & 0 & \ldots  & \mathbbm{1}_{D^{N}}
\end{array}
\right]\nonumber\\
&=&\frac{1}{2D^{N}}\left[\left(\proj{0}^{\ot N}+\proj{1}^{\ot
N}\right)\ot\mathbbm{1}_{D^{N}}\right.\nonumber\\
&&\left.+\left(\ket{0}\!\bra{1}^{\ot N}+\ket{1}\!\bra{0}^{\ot
N}\right)\ot V^{(D)}_{\pi}\right].
\end{eqnarray}
where $V^{(D)}_{\pi}$ is a permutation operator defined as
\begin{equation}
V^{(D)}_{\pi}=\sum_{i_{1},\ldots,i_{N}=0}^{D-1}\ket{i_{1}}\!\bra{i_{\pi(1)}}\ot \ket{i_{2}}\!\bra{i_{\pi(2)}}\ot\ldots
\ot\ket{i_{N}}\!\bra{i_{\pi(N)}}
\end{equation}
with $\pi$ being an arbitrary permutation of $N$--element set.
Clearly $V^{(D)}_{\pi}$ is unitary matrix for any permutation
$\pi$ and thus $\big|V_{\pi}^{(D)}\big|=\mathbbm{1}_{D^{N}}$
($|A|$ is defined as $\sqrt{A^{\dagger}A}$).
This, in view of the Lemma A.1 (Appendix), means that
$\mathcal{M}_{2}(\mathbbm{1}_{D^{N}},V_{\pi}^{(D)})\geq 0$ (for the definition
of $\mathcal{M}_{2}$ see Appendix) for any
$\pi$ and hence $\Gamma_{\mathrm{ex}}^{(2)}$ represents quantum state.
%
%
Moreover, $\Gamma_{\mathrm{ex}}^{(2)}$ may be derived from the
general form \eqref{mpbit} by substituting
$\varrho_{\mathsf{A}'}=\mathbbm{1}_{D^{N}}/D^{N}$, i.e., maximally
mixed state acting on $(\mathbb{C}^{D})^{\ot N}$. Finally, both
unitary operations in Eq. \eqref{mpbit} may be taken to be
$U_{0}=V_{\pi}^{(D)}$ and $U_{1}=\mathbbm{1}_{D^{N}}$.

As multipartite private state constitute a central notion of our
cryptographic scheme, below we shortly characterize multipartite
private states. Firstly, we notice that any state of which \textsf{c}q
state is the ideal one with respect to some basis $\mathcal{B}_{N}$
must be of the form \eqref{mpbit} and {\it vice versa}.

{\it Theorem III.1.} Let $\varrho_{\mathsf{AA}'}$ be a state
defined on $\mathcal{H}\ot\mathcal{H}'$ with
$\mathcal{H}=(\mathbb{C}^{d})^{\ot N}$ and arbitrary but finite--dimensional
$\mathcal{H}'$. Let also
$\varrho_{\mathsf{A}E}^{(\textsf{c}\mathrm{q})}$ denote the
\textsf{c}q state obtained from the purification of
$\varrho_{\mathsf{AA}'}$ upon the measurement of the $\mathsf{A}$
part in $\mathcal{B}_{N}^{\mathrm{prod}}$ and tracing out the
$\mathsf{A}'$ part. Then
$\varrho_{\mathsf{A}E}^{(\mathsf{c}\mathrm{q})}$ is of the form
\eqref{idealcq} if and only if $\varrho_{\mathsf{AA}'}$ is of the
form \eqref{mpbit}, both with respect to
$\mathcal{B}_{N}^{\mathrm{prod}}$.

This fact may be proved in exactly the same way as its bipartite
version from Ref. \cite{KH}.

Secondly, it was shown in Ref. \cite{PHRA} that any multipartite
private state is distillable providing also a lower bound on
distillable entanglement. For completeness it is desirable to
briefly recall this result, which can be stated as follows.
For any multipartite private state $\Gamma_{\mathsf{AA}'}^{(d)}$
its distillable entanglement is bounded as
\begin{eqnarray}\label{rate1}
&&E_{D}\big(\Gamma^{(d)}_{\mathsf{AA'}}\big)\geq\nonumber\\
&&\max_{\substack{i,j=0,\ldots,d-1\\i<j}}\left\{ a^{\max}_{ij}\left[1-H\left(\frac{1}{2}
+\frac{\eta_{ij}}{2\sqrt{a_{ij}^{(1)}a_{ij}^{(2)}}}\right)\right]\right\}
\nonumber\\
\end{eqnarray}
where $\eta_{ij}$, $a_{ij}^{(1)}$, $a_{ij}^{(1)}$, and finally
$a^{\max}_{ij}$ are parameters characterizing the given
private state $\Gamma_{\mathsf{AA}'}^{(d)}$. They are defined as
follows
\begin{equation}\label{mpditEta}
\eta_{ij}=\max\left|\braa{f_{1}}\ldots\braa{f_{N}}
U_{i}\varrho_{\textsf{A}'}U_{j}^{\dagger}
\kett{g_{1}}\ldots\kett{g_{N}}\right|,
\end{equation}
where maximum is taken over a pair of pure product vectors
$\kett{f_{1}}\ldots\kett{f_{N}}$ and $\kett{g_{1}}\ldots\kett{g_{N}}$
belonging to $\mathcal{H}'$.
The parameters $a^{(1)}_{ij}$ and $a^{(2)}_{ij}$ are given by
\begin{equation}\label{ampdits1}
a^{(1)}_{ij}=\bra{\widetilde{f}_{1}^{(ij)}}\ldots\bra{\widetilde{f}_{N}^{(ij)}}
U_{i}\varrho_{\textsf{A}'}U_{i}^{\dagger}\ket{\widetilde{f}_{1}^{(ij)}}\ldots
\ket{\widetilde{f}_{N}^{(ij)}}
\end{equation}
and
\begin{equation}\label{ampdits2}
a^{(2)}_{ij}=\bra{\widetilde{g}_{1}^{(ij)}}\ldots\bra{\widetilde{g}_{N}^{(ij)}}
U_{j}\varrho_{\textsf{A}'}U_{j}^{\dagger}\ket{\widetilde{g}_{1}^{(ij)}}\ldots\ket{\widetilde{g}_{N}^{(ij)}},
\end{equation}
where $ \ket{\widetilde{f}_{1}^{(ij)}}\ldots
\ket{\widetilde{f}_{N}^{(ij)}}$ and $\ket{\widetilde{g}_{1}^{(ij)}}\ldots\ket{\widetilde{g}_{N}^{(ij)}}$
are the vectors realizing the maximum in Eq. (\ref{mpditEta}).
Finally $a_{ij}^{\max}$ denotes the larger of two numbers
$a_{ij}^{(1)}$ and $a_{ij}^{(2)}$.

It follows from Eqs. \eqref{mpditEta}, \eqref{ampdits1}, and
\eqref{ampdits2} that $\eta_{ij}$ is always positive and on the
other hand $\eta_{ij}\leq \sqrt{a^{(1)}_{ij}a^{(2)}_{ij}}$. This
means that $a_{ij}^{\mathrm{max}}>0$ and consequently for any pair
$(i<j)$ the expression under the maximum in Eq. \eqref{rate1} is
positive proving that $E_{D}$ of any multipartite private state is
nonzero.
%
%

Finally, we notice following Ref. \cite{PHRA} that for bipartite
private states also other entanglement measures were bounded from
below. Namely, it was shown that
\begin{equation}
E_{C}(\gamma_{A_{1}A_{2}A_{1}'A_{2}'}^{(d)})\geq \log d
\end{equation}
and, due to the fact that entanglement of formation
is not smaller than the entanglement cost, $E_{F}(\gamma_{A_{1}A_{2}A_{1}'A_{2}'}^{(d)})\geq \log d$.
\subsection{Conditions for closeness to multipartite private states}
Here we provide necessary and sufficient conditions allowing
for judging how close to some multipartite private state is
some given state $\varrho_{\mathsf{AA}'}$ defined on $\mathcal{H}\ot\mathcal{H}'$.

Let us firstly notice that any state acting on $\mathcal{H}\ot\mathcal{H}'$
may be written in the following block form
\begin{equation}\label{form2}
\varrho_{\mathsf{AA}'}=\sum_{i_{1},\ldots,i_{N}=0}^{d-1}\sum_{j_{1},\ldots,j_{N}=0}^{d-1}
\ket{i_{1}\ldots i_{N}}\!\bra{j_{1}\ldots j_{N}}\ot\Omega_{i_{1}\ldots i_{N}}^{j_{1}\ldots j_{N}},
\end{equation}
where $\Omega_{i_{1}\ldots i_{N}}^{j_{1}\ldots j_{N}}$ are assumed
to be square matrices defined on $\mathcal{H}'$. Also by
$\widetilde{\varrho}_{\mathsf{A}}$ we denote the state
$\widetilde{\varrho}_{\mathsf{A}}=\Tr_{\mathsf{A}'}(U_{t}\varrho_{\mathsf{AA}'}U_{t}^{\dagger})$
with some twisting $U_{t}$, and by
$(\widetilde{\varrho}_{\mathsf{A}})_{i_{1}\ldots i_{N}}^{j_{1}\ldots
j_{N}}$ its entries in the standard basis. Then we can prove the
following useful lemma.

{\it Lemma III.1.} Let $\varrho_{\mathsf{AA}'}$ be some density
matrix acting on $\mathcal{H}\ot\mathcal{H}'$ with
$\mathcal{H}=(\mathbb{C}^{d})^{\ot N}$ and arbitrary
finite--dimensional $\mathcal{H}'$. Then there exists such
twisting $U_{t}$ that for a fixed index $i$ all the elements
$(\widetilde{\varrho}_{\mathsf{A}})_{i\ldots i}^{j\ldots j}$ and
$(\widetilde{\varrho}_{\mathsf{A}})_{j\ldots j}^{i\ldots i}$
$(j=0,\ldots,d-1)$ of the $i$--th row and column of
$\widetilde{\varrho}_{\mathsf{A}}=\Tr_{\mathsf{A}'}(U_{t}\varrho_{\mathsf{AA}'}U_{t}^{\dagger})$
equal $\big\|\Omega_{i\ldots i}^{j\ldots j}\big\|_{1}$ and
$\big\|\Omega_{j\ldots j}^{i\ldots i}\big\|_{1}$, respectively.

{\it Proof.} The proof is a simple extension of the one presented
in Ref. \cite{KH}. Acting on the state $\varrho_{\mathsf{AA}'}$ with an
unitary twisting $U_{t}$ and tracing out the $\mathsf{A}'$ subsystem, one gets
\begin{eqnarray}\label{rownanie}
\widetilde{\varrho}_{\mathsf{A}}&\nmss=\nmss&\sum_{i_{1},\ldots,i_{N}=0}^{d-1}\sum_{j_{1},\ldots,j_{N}=0}^{d-1}
\Tr\left(U_{i_{1}\ldots i_{N}}\Omega_{i_{1}\ldots i_{N}}^{j_{1}\ldots j_{N}}U_{j_{1}\ldots j_{N}}^{\dagger}\right)\nonumber\\
&&\hspace{3cm}\times\ket{i_{1}\ldots i_{N}}\bra{j_{1}\ldots j_{N}}.
\end{eqnarray}
First of all let us mention that from Eq. \eqref{rownanie} it
follows that we do not need to care about blocks lying on the
diagonal of $\varrho_{\mathsf{AA}'}$ as the blocks
$\Omega_{i_{1}\ldots i_{N}}^{i_{1}\ldots i_{N}}$ must be positive
and the following holds
\begin{eqnarray}
\Tr\left(U_{i_{1}\ldots i_{N}}
\Omega_{i_{1}\ldots i_{N}}^{i_{1}\ldots i_{N}}U_{i_{1}\ldots i_{N}}^{\dagger}\right)=
\norsl{\Omega_{i_{1}\ldots i_{N}}^{i_{1}\ldots i_{N}}}.
\end{eqnarray}
Now, let us focus now on the matrices $\Omega_{i\ldots i}^{j\ldots j}$
for some fixed $i$ and any $j\neq i$ (as the case of $i=j$ has just been discussed).
For simplicity and without any loss of generality
we can choose $i=0$ and thus we need to prove the theorem for
$j=1,\ldots,d-1$. At the beginning let us
concentrate on the matrix $\Omega_{0\ldots 0}^{1\ldots 1}$. We can
express it with the singular--value decomposition as
$\Omega_{0\ldots 0}^{1\ldots 1}=V_{1}D_{1}W_{1}^{\dagger}$, where
$V_{1}$ and $W_{1}$ are unitary matrices and $D_{1}$ stands for a diagonal
matrix containing singular values of $\Omega_{0\ldots 0}^{1\ldots 1}$, i.e.,
eigenvalues of $\big|\Omega_{0\ldots 0}^{1\ldots 1}\big|$. Then from
Eq. \eqref{rownanie} one infers that it suffices to take
$U_{0\ldots 0}=V_{1}^{\dagger}$ and $U_{1\ldots 1}=W$ in the twisting
$U_{t}$ to get
\begin{eqnarray}\label{rownanie2}
\Tr\left(U_{0\ldots 0}\Omega_{0\ldots 0}^{1\ldots 1}U_{1\ldots 1}^{\dagger}\right)&\nmss=\nmss&
\Tr(V^{\dagger}V D W^{\dagger}W)\nonumber\\
&\nmss=\nmss&\Tr D=\norsl{\Omega_{0\ldots 0}^{1\ldots 1}}.
\end{eqnarray}
Now we may proceed with the remaining matrices $\Omega_{0\ldots
0}^{j\ldots j}$ $(j=2,\ldots,d-1)$. We need to find such matrices
in the twisting $U_{t}$ that Eq. \eqref{rownanie2} holds also for
the remaining $\Omega_{0\ldots 0}^{j\ldots j}$. Notice that
unitary matrices $U_{0\ldots 0}$ and $U_{1\ldots 1}$ have just
been fixed, however, we have still some freedom provided by
$U_{j\ldots j}$ $(j=2,\ldots,d-1)$. Using the singular value
decomposition of all $\Omega_{0\ldots 0}^{j\ldots j}$
$(j=2,\ldots,d-1)$ we may write $\Omega_{0\ldots 0}^{j\ldots
j}=V_{j}D_{j}W_{j}^{\dagger}$. This leads to
\begin{eqnarray}\label{kolejnerownanie}
\Tr\left(U_{0\ldots 0}\Omega_{0\ldots 0}^{j\ldots j}U_{j\ldots j}^{\dagger}\right)
&\nmss=\nmss&\Tr\left(V^{\dagger}\Omega_{0\ldots 0}^{j\ldots j}U_{j\ldots j}^{\dagger}\right)\nonumber\\
&\nmss=\nmss&\Tr\left(V^{\dagger}V_{j}D_{j}W_{j}^{\dagger}U_{j\ldots j}^{\dagger}\right)\nonumber\\
&\nmss=\nmss&\Tr\left(D_{j}W_{j}^{\dagger}U_{j\ldots j}^{\dagger}V^{\dagger}V_{j}\right),
\end{eqnarray}
where we used the property of trace saying that $\Tr AB=\Tr BA$.
It is clear from the above that to get the trace norm of
$\Omega_{0\ldots 0}^{j\ldots j}$ for any $j=2,\ldots,d-1$ it
suffices to choose $U_{j\ldots j}$ in such way that
$W_{j}^{\dagger}U_{j\ldots
j}^{\dagger}V^{\dagger}_{1}V_{j}=\mathbbm{1}$. This means that
$U_{j\ldots j}=V^{\dagger}_{1}V_{j}W_{j}^{\dagger}$
$(j=2,\ldots,d-1)$. The remaining $U_{i_{1}\ldots i_{N}}$
appearing in the definition of $U_{t}$ may be chosen at will.
Concluding we showed that there exists such $U_{t}$ that for a
fixed $i$ it holds that $(\widetilde{\varrho}_{\mathsf{A}})_{i\ldots
i}^{j\ldots j}=\big\|\Omega_{i\ldots i}^{j\ldots j}\big\|_{1}$
$(j=0,\ldots,d-1)$. The fact that also
$(\widetilde{\varrho}_{\mathsf{A}})_{j\ldots j}^{i\ldots
i}=\big\|\Omega_{i\ldots i}^{j\ldots j}\big\|_{1}$ follows
obviously from hermiticity of $\widetilde{\varrho}_{\mathsf{A}}$.
$\blacksquare$

This is a very useful lemma due to the fact that twistings
do not change the \textsf{c}q state. It allows us to concentrate on a
particular form of a given state $\varrho_{\mathsf{AA}'}$. In
other words, we can think about the state $\varrho_{\mathsf{AA}'}$
as if it has such a reduction to \textsf{A} subsystem that some of
its elements in fixed row or column are trace norms of respective
blocks of $\varrho_{\mathsf{AA}'}$ (obviously with respect to the
same product basis $\mathcal{B}_{N}^{\mathrm{prod}}$). As an
illustrative example we can consider $\varrho_{\mathsf{AA}'}$ with
$d=2$. Then from Eq. \eqref{form2} it can be written as
\begin{equation}
\varrho_{\mathsf{AA}'}=\left[
\begin{array}{ccccc}
\Omega_{0\ldots 0}^{0\ldots 0} & \Omega_{0\ldots 0}^{0\ldots 1} & \ldots & \Omega_{0\ldots 0}^{1\ldots 1} \\*[1ex]
\Omega_{0\ldots 1}^{0\ldots 0} & \Omega_{0\ldots 1}^{0\ldots 1} & \ldots & \Omega_{0\ldots 1}^{1\ldots 1} \\*[1ex]
\vdots & \vdots & \ddots & \vdots \\*[1ex]
\Omega_{1\ldots 1}^{0\ldots 0} & \Omega_{1\ldots 1}^{0\ldots 1} & \ldots &  \Omega_{1\ldots 1}^{1\ldots 1}
\end{array}
\right],
\end{equation}
where $\Omega_{i_{1}\ldots i_{N}}^{j_{1}\ldots
i_{N}}=\big(\Omega_{j_{1}\ldots j_{N}}^{i_{1} \ldots
i_{N}}\big)^{\dagger}$ and $\Omega_{i_{1}\ldots
i_{N}}^{i_{1}\ldots i_{N}}\geq 0$ for any $i_{k},j_{k}=0,1$. In
view of Lemma 3.2 the above may be brought to the following
state
\begin{eqnarray}\label{twistedState}
\widetilde{\varrho}_{\mathsf{A}}&\nmss\equiv\nmss&\Tr_{\mathsf{A}'}(U_{t}\varrho_{\mathsf{AA}'}U_{t}^{\dagger})\nonumber\\
&\nmss=\nmss&
\left[
\begin{array}{ccccc}
\norsl{\Omega_{0\ldots 0}^{0\ldots 0}} &
(\widetilde{\varrho}_{\mathsf{A}})_{0\ldots 0}^{0\ldots 1} &
\ldots & \norsl{\Omega_{0\ldots 0}^{1\ldots 1}} \\*[1ex]
(\widetilde{\varrho}_{\mathsf{A}})_{0\ldots 1}^{0\ldots 0} &
\norsl{\Omega_{0\ldots 1}^{0\ldots 1}} & \ldots &
(\widetilde{\varrho}_{\mathsf{A}})_{0\ldots 1}^{1\ldots 1}
\\*[1ex] \vdots & \vdots & \ddots & \vdots \\*[1ex]
\norsl{\Omega_{1\ldots 1}^{0\ldots 0}} &
(\widetilde{\varrho}_{\mathsf{A}})_{1\ldots 1}^{0\ldots 1} &
\ldots & \norsl{\Omega_{1\ldots 1}^{1\ldots 1}}
\end{array}
\right].
\end{eqnarray}

Now we are prepared to provide the aforementioned conditions for
closeness to multipartite private states (the bipartite case was
discussed in Ref. \cite{KH}). Firstly we show that if a given
$\varrho_{\mathsf{AA}'}$ is close to some multipartite pdit then
(due to the above lemma) there exist such $U_{t}$ that the
$\mathsf{A}$ subsystem has all the elements
$(\Tr_{\mathsf{A}'}U_{t}\varrho_{\mathsf{AA}'}U_{t}^{\dagger})_{i\ldots
i}^{j\ldots j}= \big\|\Omega_{i\ldots i }^{j\ldots j}\big\|_{1}$
for $j=0,\ldots,d-1$ close to $1/d$.

{\it Theorem III.2.}  Let $\Omega_{i_{1}\ldots i_{N}}^{j_{1}\ldots
j_{N}}$ be some matrices and $\varrho_{\mathsf{AA}'}$ be an
$N$--partite state of the form (\ref{form2}) such that
$\big\|\varrho_{\mathsf{AA}'}-\Gamma_{\mathsf{AA}'}^{(d)}\big\|_{1}\leq
\epsilon$ for some multipartite private state
$\Gamma_{\mathsf{AA}'}^{(d)}$ for some $\epsilon >0$. Then for a
fixed index $i$ one has $\big|\big\|\Omega_{i\ldots i}^{j\ldots
j}\big\|_{1}-(1/d)\big|\leq \epsilon$ and
$\big|\big\|\Omega_{j\ldots j}^{i\ldots
i}\big\|_{1}-(1/d)\big|\leq \epsilon$ for any $j=0,\ldots,d-1$.

{\it Proof.} The proof is a modification of the one given in Ref.
\cite{KH} (Proposition 3). Let $\Gamma_{\mathsf{AA}'}^{(d)}$ be
such a private state that
$\big\|\varrho_{\mathsf{AA}'}-\Gamma_{\mathsf{AA}'}^{(d)}\big\|_{1}\leq
\epsilon$ and $U_{t}$ be such twisting that
$\Gamma_{\mathsf{AA}'}^{(d)}=U_{t}(P_{d,N}^{(+)}\ot\sigma_{\mathsf{A}'})U_{t}^{\dagger}$
with $\sigma_{\mathsf{A}'}$ denoting some state on $\mathcal{H}'$.
Then, due to the invariance of the trace norm under unitary
operations, we have
\begin{equation}
\left\|U_{t}^{\dagger}\varrho_{\mathsf{AA}'}U_{t}-P_{d,N}^{(+)}\ot\sigma_{\mathsf{A}'}\right\|_{1}\leq
\epsilon.
\end{equation}
Now, utilizing the fact that the trace norm can only decrease
under the partial trace,
we get
\begin{equation}
\left\|\widetilde{\varrho}_{\mathsf{A}}-P_{d,N}^{(+)}\right\|_{1}\leq
\epsilon,
\end{equation}
where $\widetilde{\varrho}_{\mathsf{A}}$ is of the form
\eqref{rownanie}. Notice that in general $U_{t}$ does not have to
be the one bringing $\widetilde{\varrho}_{\mathsf{A}}$ to the form
discussed in Lemma III.1. After application of the explicit form
of $P_{d,N}^{(+)}$ and $\widetilde{\varrho}_{\mathsf{A}}$ given by
Eq. \eqref{rownanie}, one can rewrite the above as
\begin{eqnarray}
&&\left\|\sum_{\substack{i_{1},\ldots,i_{N}=0\\j_{1},\ldots,j_{N}=0}}^{d-1}
\ket{i_{1}\ldots i_{N}}\!\bra{j_{1}\ldots
j_{N}}\Tr\left(U_{i_{1}\ldots i_{N}}\Omega_{i_{1}\ldots
i_{N}}^{j_{1}\ldots j_{N}}U_{j_{1}\ldots
j_{N}}^{\dagger}\right)\right.\nonumber\\
&&\left.\hspace{2cm}-\frac{1}{d}\sum_{i,j=0}^{d-1}\ket{i}\!\bra{j}^{\ot
N}\right\|_{1}\leq \epsilon.
\end{eqnarray}
Now we may utilize the fact that for any
$A=\sum_{ij}a_{ij}\ket{i}\bra{j}$ square of its Hilbert--Schmidt
norm is given by $\|A\|_{2}^{2}=\sum_{ij}|a_{ij}|^{2}$ and that
$\|A\|_{2}\leq \|A\|_{1}$. Therefore, if $\|A\|_{2}\leq \epsilon$
for some $\epsilon>0$ then one infers that any of its elements
obeys $|a_{ij}|\leq \epsilon$ $(i,j=0,\ldots,d-1)$. This
reasoning, after application to the matrix
$\widetilde{\varrho}_{\mathsf{A}}-P_{d,N}^{(+)}$, leads us to the
conclusion that for any $i,j=0,\ldots,d-1$ it holds
\begin{equation}
\left|\Tr\left(U_{i\ldots i}\Omega_{i\ldots i}^{j\ldots
j}U_{j\ldots j}^{\dagger}\right)-\frac{1}{d}\right|\leq \epsilon,
\end{equation}
which eventually gives $|\Tr(U_{i\ldots i}\Omega_{i\ldots
i}^{j\ldots j}U_{j\ldots j}^{\dagger})| \geq 1/d-\epsilon$. This,
after application of the polar decomposition of $\Omega_{i\ldots
i}^{j\ldots j}$ and properties of trace can be rewritten as
$|\Tr(W_{ij}|\Omega_{i\ldots i}^{j\ldots j}|)| \geq 1/d-\epsilon$
with $W_{ij}$ being some unitary matrix. Now, applying the
Cauchy--Schwarz inequality to the Hilbert--Schmidt scalar product
we can infer that for any positive $A$ and unitary $W$ the
following chain of inequalities holds
\begin{eqnarray}
\left|\Tr\left(WA\right)\right|&=&\left|\Tr\left(W\sqrt{A}\sqrt{A}\right)\right|\nonumber\\
&\leq
&\sqrt{\Tr\left(W\sqrt{A}\sqrt{A}W^{\dagger}\right)}\sqrt{\Tr
\sqrt{A}\sqrt{A}}\nonumber\\
&=&\Tr A=\|A\|_{1}.
\end{eqnarray}
Thus we have that $\|\Omega_{i\ldots i}^{j\ldots j}\|_{1}\geq
1/d-\epsilon$ for any $(i,j=0,\ldots,d-1)$.

On the other hand we can apply such twisting $\widetilde{U}_{t}$
that after application to $\varrho_{\mathsf{AA}'}$ and tracing out
the $\mathsf{A}'$ subsystem we get
$\widetilde{\varrho}_{\mathsf{A}}$ such that in its $i$th row (or
column) $(\widetilde{\varrho}_{\mathsf{A}})_{i\ldots i}^{j\ldots
j}=\|\Omega_{i\ldots i}^{j\ldots j}\|$ for $j=1,\ldots,d-1$. Then,
one easily concludes that
\begin{equation}
\left\|\widetilde{\varrho}_{\mathsf{AA}'}-\widetilde{U}_{t}^{\dagger}\Gamma_{\mathsf{AA}'}^{(d)}\widetilde{U}_{t}\right\|_{1}\leq
\epsilon
\end{equation}
with
$\widetilde{\varrho}_{\mathsf{AA}'}=\widetilde{U}_{t}\varrho_{\mathsf{AA}'}\widetilde{U}_{t}^{\dagger}$.
After the analogous reasoning as in the previous case we get
\begin{equation}
\left|\left\|\Omega_{i\ldots i}^{j\ldots
j}\right\|_{1}-\frac{1}{d}\Tr(\widetilde{W}_{ij}\sigma_{\mathsf{A}'})\right|\leq
\epsilon
\end{equation}
for some chosen $i$ and $j=0,\ldots,d-1$. Here by $W_{ij}$ we
denoted all product of the respective unitary matrices following
from product of $\widetilde{U}_{t}$ and $U_{t}$. Using the fact
that $|z_{1}-z_{2}|\geq |z_{1}|-|z_{2}|$, we infer from the above
inequality that
\begin{equation}
\left\|\Omega_{i\ldots i}^{j\ldots j}\right\|_{1}\leq
\epsilon+\frac{1}{d}\left|\Tr\left(\widetilde{W}_{ij}\sigma_{\mathsf{A}'}\right)\right|.
\end{equation}
It follows from the Cauchy--Schwarz inequality that the absolute
value on the right--hand side is not greater that one. Thus we get
the inequalities
\begin{equation}
\left\|\Omega_{i\ldots i}^{j\ldots j}\right\|_{1}\leq
\epsilon+\frac{1}{d}
\end{equation}
for the chosen $i$ and $j=0,\ldots,d-1$. Joining this facts
together we get the desired result. $\blacksquare$

Notice that in the particular case of $d=2$, discussed already in Ref. \cite{KH}
(Proposition 3), the only condition for $\big\|\Omega_{0\ldots 0}^{1\ldots 1}\big\|_{1}$
(and equivalently for $\big\|\Omega_{1\ldots 1}^{0\ldots 0}\big\|_{1}$)
is that $\big\|\Omega_{0\ldots 0}^{1\ldots 1}\big\|_{1}\geq 1/2-\epsilon$.
This is because, due to the fact that $\Tr\widetilde{\varrho}_{\mathsf{A}}=1$ and
the positivity of $\widetilde{\varrho}_{\mathsf{A}}\geq 0$
(and thus also of the $2\times 2$ matrix containing the elements
$\big\|\Omega_{i\ldots i}^{j\ldots j}\big\|_{1}$ with $i,j=0,1$)
$\big\|\Omega_{0\ldots 0}^{1\ldots 1}\big\|_{1}\leq 1/2$.



Interestingly, one may prove also a converse statement, namely if
after applying respective twisting $U_{t}$,
for some fixed row, say the $i$th one, all $\big\|\Omega_{i\ldots i}^{j\ldots j}\big\|_{1}$
are close to $1/d$, then there exists some multipartite private state
close to a given state $\varrho_{\mathsf{AA}'}$.
%

{\it  Theorem III.3.} Let $\varrho_{\mathsf{AA'}}$ given by
Eq. (\ref{form2}) be such that for a fixed $i$ the blocks $\Omega_{i\ldots i}^{j\ldots j}$ obey
$\big|\big\|\Omega_{i\ldots i}^{j\ldots j}\big\|_{1}- 1/d\big|\leq \epsilon$ for any $j=0,\ldots,d-1$ and
$0<\epsilon<1/d$.
Then there exists such a multipartite private state $\Gamma_{\mathsf{AA}'}^{(d)}$ that
\begin{eqnarray}
\norsl{\varrho_{\mathsf{AA'}}-\Gamma_{\mathsf{AA'}}^{(d)}}&\nmss\le\nmss &
\sqrt{\log2\left[2N\sqrt{d\eta(\epsilon)}\log
d+H(2\sqrt{d\eta(\epsilon)})\right]}\nonumber\\
&&+2\sqrt{d\eta(\epsilon)},
\end{eqnarray}
where $\eta(\epsilon)\to 0$ if $\epsilon\to 0$ and
consequently the function on the right--hand side tends to zero whenever $\epsilon\to 0$.
Here $H$ denotes the binary entropy.

{\it Proof.} The proof is based on the one given in Ref.
\cite{KH}. Let $U_{t}$ be such twisting that for fixed $i$ it
holds of $(\widetilde{\varrho}_{\mathsf{A}})_{i \ldots i}^{j\ldots
j}=\big\|\Omega_{i\ldots i}^{j\ldots j}\big\|_{1}$
$(j=0,\ldots,d-1)$. Then since
$(\widetilde{\varrho}_{\mathsf{A}})_{i \ldots i}^{j\ldots
j}=[(\widetilde{\varrho}_{\mathsf{A}})_{j \ldots j}^{i\ldots
i}]^{*}$ with asterisk denoting complex conjugation, the
Hilbert--Schmidt scalar product of
$\widetilde{\varrho}_{\mathsf{A}}$ and $P_{D,N}^{(+)}$ may be
expressed as
\begin{eqnarray}\label{nierownosc}
\Tr\widetilde{\varrho}_{\mathsf{A}}P^{(+)}_{d,N}&\nmss=\nmss&\frac{1}{d}\sum_{i,j=0}^{d-1}
(\widetilde{\varrho}_{\mathsf{A}})_{i\ldots i}^{j\ldots j}\nonumber\\
&\nmss=\nmss&
\frac{1}{d}\sum_{i=0}^{d-1}(\widetilde{\varrho}_{\mathsf{A}})_{i\ldots
i}^{i\ldots i}+\frac{2}{d}
\sum_{\substack{i,j=0\\i<j}}^{d-1}\mathrm{Re}(\widetilde{\varrho}_{\mathsf{A}})_{i\ldots
i}^{j\ldots j}.
\end{eqnarray}
On the other hand from the positivity of $\widetilde{\varrho}_{\mathsf{A}}$
one may prove the following inequality
\begin{equation}
\sum_{i=0}^{d-1}(\widetilde{\varrho}_{\mathsf{A}})_{i\ldots i }^{i\ldots i}\geq \frac{2}{d-1}
\sum_{\substack{i,j=0\\i<j}}^{d-1}\mathrm{Re}(\widetilde{\varrho}_{\mathsf{A}})_{i\ldots i}^{j\ldots j},
\end{equation}
which after substitution to Eq. (\ref{nierownosc}) gives
\begin{equation}\label{nierownosc2}
\Tr\widetilde{\varrho}_{\mathsf{A}}P^{(+)}_{d,N}\geq \frac{2}{d-1}
\sum_{\substack{i,j=0\\i<j}}^{d-1}\mathrm{Re}(\widetilde{\varrho}_{\mathsf{A}})_{i\ldots i}^{j\ldots j}.
\end{equation}
Now we can utilize Lemma A.2 (see Appendix) to the $d\times d$
matrix with entries $(\widetilde{\varrho}_{\mathsf{A}})_{i\ldots i
}^{j\ldots j}$ $(i,j=0,\ldots,d-1)$. Namely, due to the assumption
that for some fixed $i$ the elements
$(\widetilde{\varrho}_{\mathsf{A}})_{i\ldots i }^{j\ldots j}$
satisfy $|(\widetilde{\varrho}_{\mathsf{A}})_{i\ldots i }^{j\ldots
j}-1/d|\leq \epsilon$, we have from Lemma A.2 that
$|(\widetilde{\varrho}_{\mathsf{A}})_{i\ldots i }^{j\ldots
j}-1/d|\leq \eta(\epsilon)$ for any $i,j=0,\ldots,d-1$ with
$\eta(\epsilon)\to 0$ for $\epsilon\to 0$. This means that the
real parts of any $(\widetilde{\varrho}_{\mathsf{A}})_{i\ldots i
}^{j\ldots j}$ also satisfies the above condition. In this light we get
from Eq. \eqref{nierownosc2} that
%
%
\begin{eqnarray}\label{nierownosc3}
\Tr\widetilde{\varrho}_{\mathsf{A}}P^{(+)}_{d,N}&\nmss\geq\nmss& \frac{2}{d-1}
\sum_{\substack{i,j=0\\i<j}}^{d-1}\mathrm{Re}(\widetilde{\varrho}_{\mathsf{A}})_{i\ldots i}^{j\ldots j}\nonumber\\
&\nmss\geq \nmss& \frac{2}{d-1}\sum_{\substack{i,j=0\\i<j}}^{d-1}\left(\frac{1}{d}-\eta(\epsilon)\right)\nonumber\\
&\nmss=\nmss& \frac{2}{d-1}\frac{d(d-1)}{2}\left(\frac{1}{d}-\eta(\epsilon)\right)\nonumber\\
&\nmss=\nmss& 1-d\eta(\epsilon),
\end{eqnarray}
where the first equality follows from the fact that the respective
sum contains $d(d-1)/2$ elements. The remainder of the proof goes along the
same line as its bipartite version from Ref. \cite{KH} leading to
the claimed inequality. $\blacksquare$

Notice that to prove the theorem for the particular case of $d=2$
it suffices to assume that
$\big\|\Omega_{0\ldots 0}^{1\ldots 1}\big\|_{1}\geq 1/2-\epsilon$.

Concluding we obtained necessary and sufficient conditions for a
given state $\varrho_{\mathsf{AA}'}$ to be close to some
multipartite private state expressed in terms of the trace norm of
some blocks of $\varrho_{\mathsf{AA}'}$ (see Eq. \eqref{form2}).

\section{Distillable cryptographic key}

\subsection{Definition}
Having introduced the concept of multipartite private states we
may pass to the definition of multipartite cryptographic key. The
seminal fact behind the notion of multipartite private states is
that as shown in Refs. \cite{KH0,KH}, one can think about quantum
cryptography as a distillation of private states by means of LOCC.
In other words, we have a standard distillation scheme (as
entanglement distillation) in which we can forget about the
eavesdropper.

{\it Definition IV.1.} Let $\varrho_{\mathsf{A}}$ denote a given multipartite state acting on
$\mathbb{C}^{d_{1}}\ot\ldots\ot\mathbb{C}^{d_{N}}$ and $(\Lambda_{n})_{n=1}^{\infty}$
a sequence of LOCC operations giving $\Lambda_{n}(\varrho_{\mathsf{A}}^{\ot n})=\varrho_{\mathsf{AA}'}^{(n)}$
with $\varrho_{\mathsf{AA}'}^{(n)}$ being a state acting on $(\mathbb{C}^{d_{n}})^{\ot N}
\ot \mathcal{H}_{n}'$.
Here $\mathcal{H}_{n}'$ stands for a finite--dimensional
Hilbert space corresponding to the $\mathsf{A}'$ part of $\varrho_{\mathsf{AA}'}^{(n)}$.
Then we say that $\Lambda=(\Lambda_{n})_{n=1}^{\infty}$
is a multipartite private state distillation protocol if there exists
such a family of multipartite private states $(\Gamma_{\mathsf{AA}'}^{(d_{n})})_{n=1}^{\infty}$
that the condition
\begin{equation}
\lim_{n\to\infty}\norsl{\varrho_{\mathsf{AA}'}^{(n)}-\Gamma_{\mathsf{AA}'}^{(d_{n})}}=0
\end{equation}
holds. A rate of the protocol $\Lambda$ is defined as
$R_{\Lambda}(\varrho_{\mathsf{A}})=\limsup_{n\to \infty}[(1/n)\log d_{n}]$
and the distillable key as
\begin{equation}
K_{D}(\varrho_{\mathsf{A}})=\sup_{\Lambda}R_{\Lambda}(\varrho_{\mathsf{A}}).
\end{equation}

As shown in the bipartite case in Ref. \cite{KH}, both the
Definition II.1 and Definition IV.1 are equivalent in
the sense that if there exists LOCC protocol distilling some
multipartite private state there also exists LOPC protocol
distilling the ideal ccq state (when purification is considered)
with the same rate. As the proof from Ref. \cite{KH} may also be
applied to the multipartite case, we provide the generalized
version of the above fact below.

{\it Theorem IV.1.} The following two implications hold. Assume
that from a given state $\sigma_{\mathsf{A}}$ such that Eve has
its purification $\ket{\psi_{\mathsf{A}E}}$ one may create by LOPC
some \textsf{c}q state
$\varrho_{\mathsf{A}E}^{(\mathsf{c}\mathrm{q})}$ (see Eq.
\eqref{nccq}) obeying
$\big\|\varrho_{\mathsf{A}E}^{(\mathsf{c}\mathrm{q})}-\varrho_{\mathsf{A}E}^{(\mathrm{id})}\big\|_{1}\leq
\epsilon$ for some $\epsilon>0$ (recall that
$\varrho_{\mathsf{A}E}^{(\mathrm{id})}$ denotes the ideal
\textsf{c}q state given by Eq. \eqref{idealcq}). Then there exists
such LOCC protocol that can distill a state
$\varrho_{\mathsf{AA}'}$ from $\sigma_{\mathsf{A}}$ that satisfies
$\big\|\varrho_{\mathsf{AA}'}-\Gamma_{\mathsf{AA}'}^{(d)}\big\|_{1}\leq
2\sqrt{\epsilon}$ for some multipartite private state
$\Gamma_{\mathsf{AA}'}^{(d)}$. On the other hand if from
$\sigma_{\mathsf{A}}$ one can distill a state
$\varrho_{\mathsf{AA}'}$ close to some pdit
$\Gamma_{\mathsf{AA}'}^{(d)}$, i.e., such that
$\big\|\varrho_{\mathsf{AA}'}-\Gamma_{\mathsf{AA}'}^{(d)}\big\|_{1}\leq
\epsilon$ then there exists a LOPC protocol distilling from
$\varrho_{\mathsf{A}}$ a \textsf{c}q state such that
$\big\|\varrho_{\mathsf{A}E}^{(\mathsf{c}\mathrm{q})}-\varrho_{\mathsf{A}E}^{(\mathrm{id})}\big\|_{1}\leq
2\sqrt{\epsilon}.$ Each subsystem of the $\mathsf{A}$ part of
$\varrho_{\mathsf{A}E}^{(\mathsf{c}\mathrm{q})}$ and of the key
part of $\Gamma_{\mathsf{AA}'}^{(d)}$ is defined on
$\mathbb{C}^{d}$.

{\it Proof.} The proof goes directly along the same lines as the one
from Ref. \cite{KH}.

Interestingly, the distillable key $K_{D}$ may be used to quantify entanglement
among multipartite states. More precisely, from the definition it follows
that $K_{D}$ is monotonic under the action of LOCC operations (see e.g. \cite{MHMeasures}).
Moreover, it vanishes on multipartite states that have at least one separable cut,
which is a consequence of the straightforward multipartite generalization
of the results from Ref. \cite{Curty1,Curty2} provided in Ref. \cite{PH_przegladowka}.
Finally, as we shall show the distillable key is normalized on $GHZ$ states $P_{d,N}^{(+)}$
in the sense that $K_{D}(P_{d,N}^{(+)})=\log d$. However, firstly we need to provide
two bounds on $K_{D}$.

\subsection{Bounds on the distillable key}
The first bound is a simple multipartite generalization of
the upper bound provided in Ref. \cite{KH}, while the second bound
is a consequence of a multipartite adaptation of the Devetak--Winter
protocol \cite{DW1,DW2}. Let us start from the upper bound.

{\it Theorem IV.2.} Let $\varrho_{\mathsf{A}}$ be some $N$--partite state. Then
\begin{equation}\label{boundEntropy}
K_{D}(\varrho_{\mathsf{A}})\leq E_{r}^{\infty}(\varrho_{\mathsf{A}}),
\end{equation}
where $E_{r}^{\infty}(\varrho_{\mathsf{A}})$ is a regularized version of
the relative entropy, i.e.,
\begin{equation}
E_{r}^{\infty}(\varrho_{\mathsf{A}})=
\lim_{n\to\infty}\frac{1}{n}\inf_{\varrho^{\mathrm{sep}}_{\mathsf{A}}\in\mathcal{D}}S(\varrho_{\mathsf{A}}^{\ot n}\|\varrho^{\mathrm{sep}}_{\mathsf{A}})
\end{equation}
and $\mathcal{D}$ denotes the set of all $N$--partite fully
separable states, i.e., states of the form
\begin{equation}
\varrho_{\mathsf{A}}^{\mathrm{sep}}=\sum_{i}p_{i}\varrho_{A_{1}}^{(i)}\ot\ldots\ot
\varrho_{A_{N}}^{(i)}.
\end{equation}

{\it Proof.} The proof is a generalization of the one
from Ref. \cite{KH}.

Interestingly, we may also bound $K_{D}$ from below. For this purpose we need to
prove the following theorem.

{\it Theorem IV.3.} Let
$\varrho_{\mathsf{A}E}^{(\mathsf{c}\mathrm{q})}$ be some
multipartite $\mathsf{c}$q state acting on $(\mathbb{C}^{d})^{\ot
N}\ot\mathbb{C}^{d_{E}}$ and given by
\begin{equation}
\varrho_{\mathsf{A}E}^{(\mathrm{cq})}=\sum_{i_{1},\ldots,i_{N}=0}^{d-1}p_{i_{1}\ldots i_{N}}
\proj{i_{1}\ldots i_{N}}.
\end{equation}
Then it is arbitrarily close to the
ideal \textsf{c}q state if and only if for a chosen party $A_{i}$
all the reductions to three--partite systems $A_{i}A_{j}E$ with
$j\neq i$ are arbitrarily close to the bipartite ideal ccq state.
More precisely, if $\big\|\varrho_{\mathsf{A}E}^{(\mathsf{c}\mathrm{q})}-\varrho_{\mathsf{A}E}^{(N,\mathrm{id})}\big\|\leq\epsilon$
holds for $\epsilon>0$,
then for the fixed party $A_{i}$ the following inequalities
\begin{eqnarray}\label{Tw2tresc}
\left\|\sum_{i_{1},\ldots, i_{N}=0}^{d-1}p_{i_{1}\ldots
i_{N}}\proj{i_{i}i_{j}}\ot\varrho_{i_{1}\ldots i_{N}}^{E}
-\varrho_{\mathsf{A}E}^{(2,\mathrm{id})}\right\|_{1}\leq \epsilon\nonumber\\
\end{eqnarray}
are satisfied for $j=1,\ldots,i-1,i+1,\ldots,N$.
Conversely, assuming that for fixed $A_{i}$ the inequalities
(\ref{Tw2tresc})
%
%
hold for $\epsilon>0$ and $j\neq i$, one has
\begin{equation}
\left\|\varrho_{\mathsf{A}E}^{(\mathsf{c}\mathrm{q})}-
\varrho_{\mathsf{A}E}^{(N,\mathrm{id})}\right\|_{1}\leq(4N-3)\epsilon
\end{equation}

{\it Proof.} We proceed in two steps. In the first step
we show that if the trace norm distance between some multipartite
\textsf{c}q state $\mathsf{\varrho}_{\mathsf{A}E}^{(\mathsf{c}\mathrm{q})}$
and the ideal one is bounded by some $\epsilon>0$ then any bipartite
state arising by tracing out $N-2$ parties from the \textsf{c}q state is
close to the bipartite ideal ccq state. This part of the proof is relatively
easy since it suffices to utilize the fact that the trace norm distance
does not increase under the partial trace.
The proof of the converse statement is much more sophisticated.

Let us assume that the following
\begin{eqnarray}\label{Tw1}
\left\|\varrho_{\mathsf{A}E}^{(\mathsf{c}\mathrm{q})}-\varrho_{\mathsf{A}E}^{(N,\mathrm{id})}\right\|_{1}\leq\epsilon
\end{eqnarray}
holds for some small $\epsilon>0$. Then since the trace norm does
not increase under the partial trace we have immediately the
following set of inequalities
\begin{eqnarray}\label{Tw2}
\left\|\sum_{i_{1},\ldots, i_{N}=0}^{d-1}p_{i_{1}\ldots
i_{N}}\proj{i_{k}i_{l}}\ot\varrho_{i_{1}\ldots i_{N}}^{E}
-\varrho_{\mathsf{A}E}^{(2,\mathrm{id})}\right\|_{1}\leq \epsilon\nonumber\\
\end{eqnarray}
for any pair of indices $k,l=1,\ldots,N$. To end the first part of the proof it suffices
to substitute $\sum_{I\setminus\{k,l\}}p_{i_{1}\ldots i_{N}}\varrho_{i_{1}\ldots i_{N}}^{E}=q_{i_{l}i_{k}}\varrho_{i_{1}\ldots i_{N}}^{E}$, where summation over
$I\setminus \{k,l\}$ means that we sum over all $i_{j}$ but $i_{k}$ and $i_{l}$.

To proceed with the second part of the proof we assume that one
chosen party, say $A_{1}$, shares with the remaining $N-1$ parties
a state that is close to the bipartite ideal \textsf{c}q state. In
other words we assume that for any $j=2,\ldots,N$ the following
inequalities
\begin{eqnarray}\label{Tw3}
\left\|\sum_{i_{1},\ldots, i_{N}=0}^{d-1}p_{i_{1}\ldots
i_{N}}\proj{i_{1}i_{j}}\ot\varrho_{i_{1}\ldots i_{N}}^{E}
-\varrho_{\mathsf{A}E}^{(2,\mathrm{id})}\right\|_{1}\leq \epsilon\nonumber\\
\end{eqnarray}
are satisfied. Basing on this set of inequalities we will show
that the left--hand side of Eq. \eqref{Tw1} is bounded from above
by some linear function of $\epsilon$ vanishing for $\epsilon\to
0$. For this purpose let us denote the left--hand side of Eq.
\eqref{Tw1} by LHS and notice that it can be split into two sums
(see Eqs. \eqref{idealcq} and \eqref{nccq}), namely the one
containing the elements for $i_{1}=\ldots =i_{N}$ and the rest
ones. In this light, denoting by $I$ the set of
sequences $(i_{1},\ldots,i_{N})$ obtained by removing all those
with $i_{1}=\ldots=i_{N}$ from the set of all possible sequences,
we can write
\begin{eqnarray}\label{Tw5}
\mathrm{LHS}&=&\sum_{(i_{1},\ldots,i_{N})\in I} p_{i_{1}\ldots
i_{N}}+\sum_{i=0}^{d-1}\norsl{p_{i\ldots i}\varrho_{i\ldots
i}^{E}-\frac{1}{d}\varrho^{E}}\nonumber\\
&\nmss\leq\nmss&\sum_{(i_{1},\ldots,i_{N})\in I}
p_{i_{1}\ldots i_{N}}+\sum_{i=0}^{d-1}p_{i\ldots i}\norsl{\varrho_{i\ldots i}^{E}-\varrho^{E}}\nonumber\\
&&+\sum_{i=0}^{d-1}\left|p_{i\ldots i}-\frac{1}{d}\right|,
\end{eqnarray}
where the inequality comes from the fact that the term
$p_{i\ldots i}\varrho_{i\ldots i}^{E}$ was added and subtracted in
the second term in the first line and from the inequality $\|A+B\|_{1}\leq \|A\|_{1}+\|B\|_{1}$.
The last equality is a simple consequence of the fact that the trace norm
of any density matrix is just one. In what follows, using the inequalities
\eqref{Tw3}, we show that all the three terms appearing in the above
are bounded by linear functions of $\epsilon$ vanishing for $\epsilon\to 0$.
With this aim, utilizing once more the fact that the trace norm does not increase
under the partial trace, we can infer from Eq. \eqref{Tw3} that
\begin{equation}\label{Tw6}
\norsl{\sum_{i_{1},\ldots,i_{N}=0}^{d-1}p_{i_{1}\ldots i_{N}}\proj{i_{1}i_{j}}-\frac{1}{d}\sum_{i=0}^{d-1}\proj{ii}}\leq \epsilon
\end{equation}
for $j=2,\ldots,N$. Now we can divide all the terms appearing in the first sum into
two groups, namely, the one for $i_{1}=i_{j}$ and the remaining terms. This, after calculating
the respective norms, leads to the following inequality
\begin{eqnarray}\label{Tw8}
&&\sum_{i=0}^{d-1}\left|\sum_{i_{2},\ldots,i_{j-1},i_{j+1},\ldots,i_{N}=0}^{d-1}p_{ii_{2}
\ldots i_{j-1}ii_{j+1}\ldots i_{N}}-\frac{1}{d}\right|\nonumber\\
&&\hspace{1cm}+\sum_ {\substack{i_{1},\ldots,i_{N}=0\\i_{1}\neq
i_{j}}}^{d-1}p_{i_{1}\ldots i_{N}} \leq\epsilon.
\end{eqnarray}
Obviously, since both terms in the above are nonnegative, any of them must be
less or equal to $\epsilon$. This allows us to write the inequalities
\begin{equation}\label{Ineq47}
\sum_{i=0}^{d-1}\left|\sum_{i_{2},\ldots,i_{j-1},i_{j+1},\ldots,i_{N}=0}^{d-1}p_{ii_{2}
\ldots i_{j-1}ii_{j+1}\ldots i_{N}}-\frac{1}{d}\right|\leq \epsilon
\end{equation}
and
\begin{equation}\label{Ineq48}
\sum_ {\substack{i_{1},\ldots,i_{N}=0\\i_{1}\neq
i_{j}}}^{d-1}p_{i_{1}\ldots i_{N}} \leq\epsilon.
\end{equation}
From the sum appearing under the sign of an absolute value
in (\ref{Ineq47}) we can extract the probability $p_{i\ldots i}$, obtaining
\begin{equation}\label{Tw10}
\sum_{i=0}^{d-1}\left|p_{i \ldots i}-\frac{1}{d}+\sum_
{\substack{i_{2},\ldots
i_{j-1},i_{j+1},\ldots,i_{N}=0\\(i_{2},\ldots
i_{j-1},i_{j+1},\ldots,i_{N})\in\mathcal{I}_{i}}}^{d-1}
p_{ii_{2}\ldots i_{j-1}i i_{j+1} \ldots i_{N}}\right| \leq
\epsilon,
\end{equation}
where $\mathcal{I}_{i}$ denotes the strings of $N-2$ indices
$(i_{2},\ldots,i_{j-1},i_{j+1},\ldots,i_{N})$ such
that at least one of them is different from $i$.
Utilizing a simple inequality $|z_{1}-z_{2}|\geq |z_{1}|-|z_{2}|$ satisfied
by all $z_{1},z_{2}\in\mathbb{C}$, we get
\begin{eqnarray}\label{Tw11}
\sum_{i=0}^{d-1}\left|p_{i \ldots i}-\frac{1}{d}\right|&\nmss\leq
\nmss&\epsilon+
\sum_{i=0}^{d-1}\sum_{(i_{2},\ldots,i_{j-1},i_{j+1},\ldots,i_{N})\in\mathcal{I}_{i}}
\hspace{-1cm}p_{ii_{2}\ldots i_{j-1}i i_{j+1}\ldots i_{N}}\nonumber\\
&\nmss= \nmss &\epsilon+
\sum_{(i_{1},\ldots,i_{N})\in\widetilde{\mathcal{I}}_{j}}
p_{i_{1}\ldots \ldots i_{N}},
\end{eqnarray}
where $\widetilde{\mathcal{I}}_{j}$ denotes the string of $N$
indices such that the first and $j$th ones are equal ($i_{1}=
i_{j}$) and at least one of the remaining ones is different from
$i_{1}$. One sees that the second term on the right--hand side of
Eq. \eqref{Tw11} may be bounded from above in the following way
\begin{eqnarray}\label{Tw13}
\sum_{(i_{1},\ldots,i_{N})\in\widetilde{\mathcal{I}}_{j}}
p_{i_{1}\ldots i_{N}}&\nmss\leq\nmss& \sum_{k=2}^{j-1}\sum_{\substack{i_{1},\ldots,i_{N}=0\\i_{k}\neq i_{1}}}^{d-1}p_{i_{1}\ldots i_{N}}\nonumber\\
&&+\sum_{k=j+1}^{N}\sum_{\substack{i_{1},\ldots,i_{N}=0\\i_{k}\neq i_{1}}}^{d-1}p_{i_{1}\ldots i_{N}}\nonumber\\
&\nmss\leq\nmss &\sum_{k=2}^{j-1}\epsilon+\sum_{k=j+1}^{N}\epsilon\nonumber\\
&\nmss=\nmss& (N-2)\epsilon,
\end{eqnarray}
where the second inequality is a consequence of the inequality
given in Eq. \eqref{Ineq48}. Finally, application of Eq. \eqref{Tw13}
to Eq. \eqref{Tw11}, gives
\begin{equation}\label{Tw14}
\sum_{i=0}^{d-1}\left|p_{i\ldots i}-\frac{1}{d}\right|\leq (N-1)\epsilon.
\end{equation}
This is a quite natural conclusion saying that if the measurement
outcomes between fixed party (here $A_{1}$) and each of the
remaining ones are almost perfectly correlated then the
measurement outcomes are almost perfectly correlated among all the
parties.

We have still two terms in Eq. \eqref{Tw5} unbounded.
Using once again the inequality $|z_{1}-z_{2}|\geq |z_{1}|-|z_{2}|$ $(z_{1},z_{2}\in\mathbb{C})$
and the fact that $p_{i_{1}\ldots i_{N}}$ represents some probability distribution
we may write
\begin{eqnarray}\label{Tw17}
\sum_{(i_{1},\ldots,i_{N})\in I}
p_{i_{1}\ldots i_{N}}&\nmss=\nmss&1-\sum_{i=0}^{d-1}p_{i\ldots i}\nonumber\\
&\nmss\leq \nmss&1-[1-(N-1)\epsilon]\nonumber\\
&\nmss=\nmss&(N-1)\epsilon.
\end{eqnarray}
Thus, the only thing we need is to bound from above the last term
in Eq. \eqref{Tw5}. Remarkably, to achieve this aim it suffices to
utilize a single inequality from the whole set \eqref{Tw3}, say
the one for $j=2$. The we can write
\begin{widetext}
\begin{eqnarray}\label{Tw19}
&&\norsl{\sum_{i_{1},\ldots,i_{N}=0}^{d-1}p_{i_{1}\ldots i_{N}}\proj{i_{1}i_{2}}\ot
\varrho_{i_{1}\ldots i_{N}}^{E}-\frac{1}{d}\sum_{i=0}^{d-1}\proj{ii}\ot\varrho^{E}}\nonumber\\
&&=\left\|\sum_{i=0}^{d-1}p_{i\ldots
i}\proj{ii}\ot\varrho_{i\ldots
i}^{E}-\frac{1}{d}\sum_{i=0}^{d-1}\proj{ii}\ot\varrho^{E}+\hspace{-0.4cm}
\sum_{(i_{1},\ldots,i_{N})\in\widetilde{\mathcal{I}}_{2}}
p_{i_{1}\ldots i_{N}}\proj{i_{1}i_{1}}\ot\varrho_{i_{1}\ldots
i_{N}}^{E} +
\hspace{-0.4cm}\sum_{\substack{i_{1},\ldots,i_{N}=0\\i_{1}\neq i_{2}}}^{d-1}
\hspace{-0.4cm}p_{i_{1}\ldots i_{N}}\proj{i_{1}i_{2}}\ot\varrho_{i_{1}\ldots i_{N}}^{E}\right\|_{1}.\nonumber\\
\end{eqnarray}
Then, due to the fact that $\|A-B\|_{1}\geq \|A\|_{1}-\|B\|_{1}$, we may rewrite the above
as
\begin{eqnarray}\label{Tw20}
\norsl{\sum_{i=0}^{d-1}p_{i\ldots i}\proj{ii}\ot\varrho_{i\ldots
i}^{E}-\frac{1}{d}\sum_{i=0}^{d-1}\proj{ii}\ot\varrho^{E}}&\nmss\leq\nmss&
\epsilon
+\sum_{(i_{1},\ldots,i_{N})\in\widetilde{\mathcal{I}}_{2}}p_{i_{1}\ldots
i_{N}}
+\sum_{\substack{i_{1},\ldots,i_{N}=0\\i_{1}\neq i_{2}}}^{d-1}p_{i_{1}\ldots i_{N}}\nonumber\\
&\nmss\leq\nmss & \epsilon+(N-2)\epsilon+\epsilon=N\epsilon,
\end{eqnarray}
where the second inequality follows from Eqs. \eqref{Tw8} and \eqref{Tw13} (with $j=2$).
On the other hand, we can easily show that
\begin{eqnarray}\label{Tw21}
\norsl{\sum_{i=0}^{d-1}p_{i\ldots i}\proj{ii}\ot\varrho_{i\ldots
i}^{E}-\frac{1}{d}\sum_{i=0}^{d-1}\proj{ii}\ot\varrho^{E}} \geq
\sum_{i=0}^{d-1}p_{i\ldots i}\norsl{\varrho_{i\ldots
i}^{E}-\varrho^{E}}-
\sum_{i=0}^{d-1}\left|p_{i\ldots i}-\frac{1}{d}\right|.\nonumber\\
\end{eqnarray}
Comparison of Eqs. \eqref{Tw14}, \eqref{Tw20} and \eqref{Tw21} allows us to write
\begin{eqnarray}\label{Tw22}
\sum_{i=0}^{d-1}p_{i\ldots i}\norsl{\varrho_{i\ldots i}^{E}-\varrho^{E}}&\nmsss\leq \nmsss& N\epsilon+
\sum_{i=0}^{d-1}\left|p_{i\ldots i}-\frac{1}{d}\right|\nonumber\\
&\nms\nms\leq\nms\nms& N\epsilon+(N-1)\epsilon\nonumber\\
&\nms\nms=\nms\nms&(2N-1)\epsilon.
\end{eqnarray}
Putting now all the pieces together, that is, substituting Eqs. \eqref{Tw14}, \eqref{Tw17}, and \eqref{Tw22}
to Eq. \eqref{Tw5}, we finally have
\begin{equation}
\norsl{\sum_{i_{1},\ldots, i_{N}=0}^{d-1}p_{i_{1}\ldots
i_{N}}\proj{i_{1}\ldots i_{N}} \ot\varrho_{i_{1}\ldots i_{N}}^{E}-
\frac{1}{d}\sum_{i=0}^{d-1}\proj{i}^{\ot N}\ot\varrho^{E}}\leq
(4N-3)\epsilon.
\end{equation}
\end{widetext}
Noting that for fixed $N$ it holds that $(4N-3)\epsilon\to 0$
whenever $\epsilon\to 0$ we finish the proof. $\blacksquare$

It should be mentioned that as it follows from the second part of
the proof, we do not need to assume the whole set of inequalities
given in \eqref{Tw3}. Actually it suffices to assume that a single
inequality from the set \eqref{Tw3} holds and the remaining ones
from the set given in Eq. \eqref{Tw6}. In other words it suffices
to assume that the measurement outcomes between a fixed party and
any from the other parties are almost perfectly correlated and
that Eve is almost completely uncorrelated from the measurement
outcomes of a single pair. This is in full agreement with our
intuition. Namely, if the measurement outcomes of any pair
$A_{i}A_{j}$ (with fixed $i$ and arbitrary $j\neq i$) are
perfectly correlated and Eve has a full knowledge about the
measurement outcomes of just a single party, she actually has the
knowledge about measurement outcomes of all parties. Therefore if
all the parties have perfect correlations and Eve is completely
uncorrelated from a single party, she must be completely
uncorrelated from all the parties. Consequently, it is sufficient
to assume that a single pair shares state that is close to a ccq
state and other chosen pairs have almost perfect correlations.

Now we are prepared to provide a lower bound on the multipartite
distillable key in the LOPC paradigm. We achieve this by extending
of the Devetak--Winter protocol to the multipartite case. We do
this by applying the bipartite Devetak--Winter protocol to $N-1$
pairs of parties in some state $\varrho_{\mathsf{A}E}$ such that
each of them consist of one chosen party, say $A_{1}$, and one of
the remaining ones. Everything works as in the standard
Devetak--Winter protocol, i.e., the party $A_{1}$ performs the
measurement in some basis, e.g. the standard one obtaining the
so--called c\textsf{q} state
(classical--quantum--$\ldots$--quantum)
\begin{equation}\label{generalcq}
\varrho_{\mathsf{A}E}^{(c\textsf{q})}=\sum_{i}p_{i}\proj{i}_{A_{1}}\ot \varrho_{A_{2}\ldots A_{N}E}^{(i)}.
\end{equation}
Then, roughly speaking, the party $A_{1}$ performs the
Devetak--Winter protocol with the remaining parties
simultaneously. One knows that the correlation between $A_{1}$ and
the remaining parties $A_{j}$ $(j=2,\ldots,N)$ are described by
the mutual information
$I(A_{1}\!:\!A_{j})(\varrho_{A_{1}A_{j}E}^{(\mathrm{cqq})})$.
However, the establish common multipartite key we need to consider
the worst case, i.e.,
$\min_{j=2,\ldots,N}I(A_{1}\!:\!A_{j})(\varrho_{A_{1}A_{j}E}^{(\mathrm{cqq})})$.
On the other hand, the correlation between $A_{1}$ and $E$ are
given by $I(A_{1}\!:\!E)$ and this amount of bits has to be
substracted at the privacy amplification stage of the process.
%
%
%

Consequently, the rate of the protocol is
\begin{equation}\label{rate}
\min_{j=2,\ldots,N}I(A_{1}\!:\!A_{j})(\varrho_{A_{1}A_{j}E}^{(\mathrm{cqq})})-
I(A_{1}\!:\!E)(\varrho_{A_{1}A_{j}E}^{(\mathrm{cqq})})
\end{equation}
and therefore, the multipartite distillable key satisfies
\begin{equation}\label{DWbound1}
C_{D}(\varrho_{\mathsf{A}E})\geq
\min_{j=2,\ldots,N}I(A_{1}\!:\!A_{j})(\varrho_{A_{1}A_{j}E}^{(\mathrm{cqq})})-
I(A_{1}\!:\!E)(\varrho_{A_{1}A_{j}E}^{(\mathrm{cqq})}).
\end{equation}
Here, $\varrho_{A_{1}A_{j}E}^{(\mathrm{cqq})}$ denotes the cqq
state, which arises from \eqref{generalcq} by tracing out all the
parties but the first and $j$th one and Eve. Moreover, by
$I(X\!:\!Y)(\varrho_{XY})$ we denoted the mutual information
defined as
$I(X\!:\!Y)(\varrho_{XY})=S(\varrho_{X})+S(\varrho_{Y})-S(\varrho_{XY})$
with $S$ denoting the von Neumann entropy.

We have still some freedom in choosing the distributing party and
therefore we can always choose the one for which the rate of the
extended Devetak--Winter protocol is highest. In this way we get
the lower bound on $C_{D}$ of the form
\begin{eqnarray}\label{DWbound2}
&&C_{D}(\varrho_{\mathsf{A}E})\geq
\max_{i=1,\ldots,N}\nonumber\\
&&\times\left[\min_{\substack{j=1,\ldots,N\\j\neq
i}}I(A_{i}\!:\!A_{j})(\varrho_{A_{i}A_{j}E}^{(\mathrm{cqq})})
-I(A_{i}\!:\!E)(\varrho_{A_{i}A_{j}E}^{(\mathrm{cqq})})\right],\nonumber\\
\end{eqnarray}

Let us finally mention that due to the Theorem IV.1 we can also
bound $K_{D}$ from below using \eqref{DWbound2}. Namely, since
$K_{D}(\varrho_{\mathsf{A}})=C_{D}(\ket{\psi_{\mathsf{A}E}})$, we
have the following
\begin{equation}\label{DWbound3}
K_{D}(\varrho_{\mathsf{A}})\geq
\max_{i=1,\ldots,N}\left[\min_{\substack{j=1,\ldots,N\\j\neq i}}I(A_{i}\!:\!A_{j})-
I(A_{i}\!:\!E)\right],
\end{equation}
where the respective quantities are calculated from e.g., the \textsf{c}q state
following the purification of $\varrho_{\mathsf{A}}$.

Now we can go back to the definition of $K_{D}$. As previously
mentioned, it holds that $K_{D}(P_{d,N}^{(+)})=\log d$. To show it
explicitly, on the one hand we can utilize the above bound. We
know from Theorem IV.1 that
$K_{D}(P_{d,N}^{(+)})=C_{D}(\ket{\psi_{\mathsf{A}E}^{(+)}})$,
where $\ket{\psi_{\mathsf{A}E}^{(+)}}$ is a purification of
$P_{d,N}^{(+)}$ and obviously has the form
$\ket{\psi_{d,N}^{(+)}}_{\mathsf{A}}\ket{E}$ with $\ket{E}$ being
some state kept by Eve. Measurement of the $\mathsf{A}$ subsystem of
$\ket{\psi_{\mathsf{A}E}^{(+)}}$ with respect to the standard
basis leads us to the ideal \textsf{c}q state
$\varrho_{\mathsf{A}E}^{(\textsf{c}\mathrm{q})}=\omega_{\mathsf{A}}^{(d,N)}\ot\proj{E},$
where
\begin{equation}
\omega_{\mathsf{A}}^{(d,N)}=\frac{1}{d}\sum_{i=0}^{d-1}\proj{i}^{\ot N},
\end{equation}
which has the quantities $I(A_{i}\!:\!A_{j})=\log d$ $(i,j=1,\ldots,N)$
and $I(A_{i}\!:\!E)=0$ $(i=1,\ldots,N)$. Substituting both these facts into Eq.
\eqref{DWbound2} gives us $K_{D}(P_{d,N}^{(+)})\geq \log d$.

On the other hand we can utilize the bound given in Eq. \eqref{boundEntropy}.
Firstly, notice that $S(\rho^{\ot n}\|\sigma^{\ot n})=nS(\rho\|\sigma)$ for two
an arbitrary natural number $n$ and arbitrary density matrices $\rho$ and $\sigma$.
Secondly, one easily finds that (see e.g. Ref. \cite{RelativeEntanglement})
\begin{equation}
S\big(P_{d,N}^{(+)}\big\|\omega_{\mathsf{A}}^{(d,N)}\big)=\log d
\end{equation}
and consequently the following estimate holds
\begin{eqnarray}
K_{D}(P_{d,N}^{(+)})&\nmss=\nmss&\lim_{n\to\infty}\frac{1}{n}
\inf_{\varrho_{\mathsf{A}}^{\mathrm{sep}}\in\mathcal{D}}S(P_{d,N}^{(+)\ot n}\|\varrho_{\mathsf{A}}^{\mathrm{sep}})\nonumber\\
&\nmss\leq\nmss&\lim_{n\to\infty}\frac{1}{n}S\big(P_{d,N}^{(+)\ot n}\big\|\omega_{\mathsf{A}}^{(d,N)\ot n}\big)\nonumber\\
&\nmss=\nmss&\lim_{n\to\infty}\frac{1}{n}nS\big(P_{d,N}^{(+)}\big\|\omega_{\mathsf{A}}^{(d,N)}\big)\nonumber\\
&\nmss=\nmss&\log d.
\end{eqnarray}
Thus $K_{D}(P_{d,N}^{(+)})\leq \log d$ and taking into account the previously
obtained inequality $K_{D}(P_{d,N}^{(+)})\leq \log d$ we infer $K_{D}(P_{d,N}^{(+)})=\log d$.
Thus, as stated previously, the multipartite distillable key may be
considered as a entanglement measure.

Let us discuss the last issue of this section.
To apply the extended Devetak--Winter protocol
successfully, that is to get a nonzero rate, one obviously has to have
the right--hand side of Eq. \eqref{DWbound2} positive.
One knows from Theorem IV.1 that distillation of some multipartite
private state by means of LOCC is equivalent to the distillation
of an ideal \textsf{c}q state by means of LOPC. This in turn means
that the closer some particular state $\varrho_{\mathsf{AA}'}$ is
to some multipartite private state, the closer is a \textsf{c}q
state obtained from it to the ideal \textsf{c}q state.
Then, from Theorem IV.3 it follows that the closer some
\textsf{c}q state is to the ideal \textsf{c}q state the closer are
its bipartite reductions to the bipartite ideal ccq state. Both
these facts mean that by distilling some multipartite private
state from copies of a given state we can make the right--hand
side of Eq. \eqref{DWbound2} (equivalently Eq. \eqref{DWbound3})
positive. Consequently, concatenating some LOCC protocol
distilling multipartite private states (an example of such a
protocol is given in the following subsection) and the extended
Devetak--Winter protocol introduces a subtle effect here. Namely,
on the one hand, using more copies in the LOCC protocol producing
a state that is closer to some multipartite private state makes
the right--hand side of Eq. \eqref{DWbound2} larger. On the other
hand spending more copies decreases the success probability which
needs to be included in the overall rate of the protocol. This
issue will become more clear when some particular classes of
states will be investigated in the next section.

\subsection{Recursive LOCC protocol distilling multipartite private states}
\label{LOCCProtocol}
Here we provide an illustrative example of a recursive LOCC
protocol allowing for distillation of multipartite private states
from some classes of states. This protocol is a generalization of
the LOCC protocol discussed in Ref. \cite{KH} to the case of an
arbitrary number of parties. Assume then that $N$ parties
$A_{1},\ldots,A_{N}$ have $k$ copies of some state
$\rho_{\mathsf{AA}'}$ in their possession. In $i$th step each
party performs the following operations.
\begin{itemize}
    \item Take the state $\rho_{\mathsf{AA'}}^{(i-1)}$ (where $\rho_{\mathsf{AA'}}^{(0)}=\rho_{\mathsf{AA}'}$)
    and one of the remaining $k-i$ copies of $\rho_{\mathsf{AA'}}$.
    \item Treating $\mathsf{A}$ part of
    $\rho_{\mathsf{AA'}}^{(i-1)}$ ($\rho_{\mathsf{AA'}}$) as source (target)
    qubits, perform CNOT operations.

    \item Finally, the parties perform the measurement in computational
    basis on the target qubits and compare the results:
    in the case of equal results (all zeros or all ones) the parties
    keep the state, otherwise they get rid of it.

\end{itemize}
In this way, spending $k$ copies of some state
$\rho_{\mathsf{AA}'}$, all the parties can distill a state
$\varrho_{\mathsf{AA}'}^{(k)}$ that is closer to some multipartite
private state than the initial one, i.e., $\rho_{\mathsf{AA}'}$.
Quantitative analysis concerning this protocol after application
to two different constructions of states may be found in Sections
\ref{Giechazety} and \ref{Giechazety2}.

\subsection{Multipartite privacy squeezing}
\label{PrivacySqueezing}

Concluding the discussion concerning the distillable key we need
to mention the multipartite version of the so--called {\it privacy
squeezing} \cite{KH0,KH} together with its application in the recent
important method \cite{HPHH} of bounding the secret key from
below. Following Lemma III.1 we know that having some state
$\varrho_{\mathsf{AA}'}$ expressed in the form \eqref{form2},
there always exists a twisting $U_{t}$ that the state
$\widetilde{\varrho}_{\mathsf{A}}=\Tr_{\mathsf{A}'}(U_{t}\varrho_{\mathsf{AA}'}U_{t}^{\dagger})$
has some special form. Namely, in some chosen row (column) some of
its entries are trace norms of respective blocks of
$\varrho_{\mathsf{AA}'}$. We will
call the state $\widetilde{\varrho}_{\mathsf{A}}$ obtained
in this way {\it privacy squeezed state}. Furthermore, we already know
that twistings do not change the $\mathsf{c}$q state with respect
to some basis $\mathcal{B}_{N}^{\mathrm{prod}}$.

Let us now proceed by stating some of the conclusion
following both the above facts. As previously mentioned we have that
$K_{D}(\varrho_{\mathsf{A}})=C_{D}(\ket{\psi_{\mathsf{A}E}})$, where
$\ket{\psi_{\mathsf{A}E}}$ denotes the purification of $\varrho_{\mathsf{A}}$.
Assuming that all the parties share some state $\varrho_{\mathsf{AA}'}$ defined on
$\mathcal{H}\ot\mathcal{H}'$ and denoting by $\ket{\psi_{\mathsf{AA}'E}}$ the purification
of $\varrho_{\mathsf{AA}'}$, we have
\begin{equation}\label{szacowanieKluczy}
K_{D}(\varrho_{\mathsf{AA}'})=C_{D}(\ket{\psi_{\mathsf{AA}'E}})\geq C_{D}(\varrho_{\mathsf{A}E})\geq
C_{D}(\varrho_{\mathsf{A}E}^{(\mathsf{c}\mathrm{q})}).
\end{equation}
Here
$\varrho_{\mathsf{A}E}=\Tr_{\mathsf{A}'}\proj{\psi_{\mathsf{AA}'E}}$
and $\varrho_{\mathsf{A}E}^{(\mathsf{c}\mathrm{q})}$ stands for a
\textsf{c}q state obtained upon the measurement of the
$\mathsf{A}$ subsystem in $\mathcal{B}_{N}^{\mathrm{prod}}$. The
first inequality follows from the fact that throwing out the
$\mathsf{A}'$ subsystem one can only lower the key as it could be
treated 'virtually' as giving it to Eve. The second inequality is
a consequence of the fact that measurement in some product basis
leads to classical state on the $\mathsf{A}$ part of the state
(notice that such measurement is LOPC operation which due to the
definition of $C_{D}$ can only lower its value).

Now we can formulate and prove the following theorem as a
multipartite generalization of the bipartite considerations from
Ref. \cite{HPHH} (cf. \cite{KHPhD}) which exploit privacy
squeezing to bound the secure key from below.

{\it Theorem IV.4.} Let $\varrho_{\mathsf{AA}'}$ be some $N$--partite
state defined on $\mathcal{H}\ot\mathcal{H}'$. Then
\begin{equation}\label{theoremIV4}
K_{D}(\varrho_{\mathsf{AA}'})\geq
C_{D}(\widetilde{\varrho}_{\mathsf{A}E}^{(\mathsf{c}\mathrm{q})}),
\end{equation}
where $\widetilde{\varrho}_{\mathsf{A}E}^{(\mathsf{c}\mathrm{q})}$ is a \textsf{c}q state
derived from purification $\ket{\widetilde{\varrho}_{\mathsf{A}E}}$ of privacy squeezed state
$\widetilde{\varrho}_{\mathsf{A}}=\Tr_{\mathsf{A}'}(U_{t}\varrho_{\mathsf{AA}'}U_{t}^{\dagger})$.

{\it Proof.} Denoting by $\ket{\psi_{\mathsf{AA}'E}}$ the purification of
$\varrho_{\mathsf{AA}'}$, we have immediately from Eq. \eqref{szacowanieKluczy}
that $K_{D}(\varrho_{\mathsf{AA}'})
\geq C_{D}(\varrho_{\mathsf{A}E}^{(\mathsf{c}\mathrm{q})})$ with
$\varrho_{\mathsf{A}E}^{(\mathsf{c}\mathrm{q})}$ standing for a
$\mathsf{c}$q state being a result of the measurement of
$\mathsf{A}$ part in $\mathcal{B}_{N}^{\mathrm{prod}}$ and tracing
$\mathsf{A}'$ part of $\ket{\psi_{\mathsf{AA}'E}}$. Then, as
already stated, for any twisting $U_{t}$ (in
$\mathcal{B}_{N}^{\mathrm{prod}}$) the states
$\varrho_{\mathsf{AA}'}$ and
$\widetilde{\varrho}_{\mathsf{AA}'}\equiv
U_{t}\varrho_{\mathsf{AA}'}U_{t}^{\dagger}$ have the same
\textsf{c}q states with respect to the basis
$\mathcal{B}_{N}^{\mathrm{prod}}$. Consequently,
$C_{D}(\varrho_{\mathsf{A}E}^{(\mathsf{c}\mathrm{q})})=
C_{D}(\sigma_{\mathsf{A}E}^{(\mathsf{c}\mathrm{q})})$ with
$\sigma_{\mathsf{A}E}^{(\mathsf{c}\mathrm{q})}$ being a
\textsf{c}q state derived from the twisted state
$U_{t}\varrho_{\mathsf{AA}'}U_{t}^{\dagger}$ (obviously {\it via}
its purification). Now, we can consider the situation in which the
$\mathsf{A}'$ subsystem is given to Eve. This means that instead
of taking 'huge' purification
$\ket{\widetilde{\psi}_{\mathsf{AA'}E}}$ of the privacy squeezed
state
$\widetilde{\varrho}_{\mathsf{A}}=\Tr_{\mathsf{A}'}\widetilde{\varrho}_{\mathsf{AA}'}
=\Tr_{\mathsf{A}'}(U_{t}\varrho_{\mathsf{AA}'}U_{t}^{\dagger})$ we
can take a 'smaller' version denoted by
$\ket{\widetilde{\varrho}_{\mathsf{A}E}}$ (more precisely to
purify some density matrix acting on $\mathcal{H}$ it suffices to
use a Hilbert space of lower dimensionality than to purify a state
acting on $\mathcal{H}\ot\mathcal{H}'$). The new purification
obviously must obey
$\widetilde{\varrho}_{\mathsf{A}}=\Tr_{E}\proj{\widetilde{\varrho}_{\mathsf{A}E}}$.
Now comparing these two situations we infer that
$C_{D}(\sigma_{\mathsf{A}E}^{(\mathsf{c}\mathrm{q})})\geq
C_{D}(\widetilde{\varrho}_{\mathsf{A}E}^{(\mathsf{c}\mathrm{q})})$
holds, where
$\widetilde{\varrho}_{\mathsf{A}E}^{(\mathsf{c}\mathrm{q})}$ is
$\mathsf{c}$q state appearing upon measurement of $\mathsf{A}$
subsystem of $\ket{\widetilde{\varrho}_{\mathsf{A}E}}$ in
$\mathcal{B}_{N}^{\mathrm{prod}}$. The inequality is a consequence
of the fact that in the case of the first \textsf{c}q state the
$\mathsf{A}'$ part unused, however, kept by the parties. In turn,
in the second situation the $\mathsf{A}'$ subsystem is treated as
it would be given to Eve when deriving
$\widetilde{\varrho}_{\mathsf{A}E}^{(\mathsf{c}\mathrm{q})}$.
Giving some part of state can only lower the secrecy as in this
case, roughly speaking, she gains some information about what is
shared by the parties. This concludes the proof. $\blacksquare$

\section{Constructions}
In this section we present two constructions of multipartite bound
entangled states with nonzero distillable cryptographic key. Both
are based on the structure exhibited by the GHZ states and
therefore the scheme of secure key distillation presented above
easily applies here.

The first construction is a straightforward generalization of the
bipartite construction presented in Ref. \cite{PHRANATO}.
Therefore, for comparative purposes, we present also a plot
containing a lower bound on distillable key in the bipartite case.
The second construction is completely new and in comparison to the
first one allows to get a higher lower bounds on distillable key
than the first one.

Before we start it is desirable to establish the notation that we
will use extensively below. By $\mathcal{P}_{0}^{(N)}$ we shall
denote a projector onto the $N$--partite pure state
$\ket{\psi_{0}^{(N)}}=\ket{0}^{\ot N}$ and $\mathcal{P}_{i}^{(N)}$
$(i=1,\ldots,N)$ is a projector onto the $N$--partite state
$\ket{\psi_{i}^{(N)}}$, in which the $i$th party possesses
$\ket{1}$, while other particles are in the $\ket{0}$ state. For
instance $\mathcal{P}_{2}^{(4)}$ denotes the projector onto the
four--partite pure state $\ket{\psi_{2}^{(4)}}=\ket{0100}$.
Moreover, let $\overline{\mathcal{P}}_{0}^{(N)}$ and
$\overline{\mathcal{P}}_{i}^{(N)}$ denote projectors obtained from
$\mathcal{P}_{0}^{(N)}$ and $\mathcal{P}_{i}^{(N)}$, respectively,
by exchanging all zeros and ones. Thus, for example
$\overline{\mathcal{P}}_{2}^{(4)}$ is the projector onto
$\ket{\overline{\psi}_{2}^{(4)}}=\ket{1011}$. We will denote in an analogous way
by $\ket{\psi_{ij}^{(N)}}$ ($\ket{\overline{\psi}_{ij}^{(N)}}$)
a $N$--qubit pure state, in which $i$th and $j$th
qubits are in the $\ket{1}$ ($\ket{0}$) state and the remaining ones are
in the $\ket{0}$ ($\ket{1}$) state. Then by
$\mathcal{P}_{ij}^{(N)}$ and $\overline{\mathcal{P}}_{ij}^{(N)}$
we denote projectors onto $\ket{\psi_{ij}^{(N)}}$ and
$\ket{\overline{\psi}_{ij}^{(N)}}$, respectively.

Let also $T_{i}$ denote the partial transposition with respect to
$i$th party (with the exception that $T_{0}$ denotes the identity
map). Here we usually assume that each party has two subsystems of
a given state $\varrho_{\mathsf{AA}'}$ and sometimes $T_{i}$ will
be denoting the partial transposition with respect to one or both
subsystems. It will be, however, clear from the context which of
the subsystems are partially transposed. Concatenation of partial
transpositions with respect to some subset of parties, say
$A_{1},\ldots,A_{k}$ will be denoted by $T_{1,\ldots,k}$.

%

%
%

%
\subsection{The first construction}
\label{Giechazety}
Here we assume that the key part on each site is of qubit
structure, while the shield part has arbitrary dimension,
however, with the same dimension on each site. More precisely, we have
$\mathcal{H}_{i}=\mathbb{C}^{2}$ and
$\mathcal{H}_{i}'=\mathbb{C}^{D}$ $(i=1,\ldots,N)$.

Now, let us introduce the following matrix
\begin{equation}\label{XDn}
X_{D}^{(N)}=\frac{1}{D^{N}+2D-4}\left[(D-2)P^{(+)}_{D,N}-2P_{D}^{(N)}+Q_{D}^{(N)}\right],
\end{equation}
where, as previously, $P^{(+)}_{D,N}$ denotes a projector onto the
$N$--partite $D$--dimensional GHZ state (see Eq.
\eqref{GHZstates}), and $P_{D}^{(N)}$ and $Q_{D}^{(N)}$ are
projectors defined as
\begin{equation}\label{projektory1}
P_{D}^{(N)}=R_{D}^{(N)}-P^{(+)}_{D,N}, \qquad
Q_{D}^{(N)}=\mathbbm{1}_{D^{N}}-R_{D}^{(N)},
\end{equation}
where
\begin{equation}\label{projektory2}
R_{D}^{(N)}=\sum_{i=0}^{D-1}\proj{i}^{\ot N}.
\end{equation}
The projectors $P_{D}^{(N)}$ and $Q_{D}^{(N)}$ are chosen in such
a way that each operator from the triple $P^{(+)}_{D,N}$,
$P_{D}^{(N)}$, and $Q_{D}^{(N)}$ is defined on orthogonal support.
Furthermore, the denominator in Eq. \eqref{XDn} is chosen such
that the matrix $X_{D}^{(N)}$ is normalized in the trace norm.

The states under consideration are of the form
\begin{eqnarray}\label{Constr1}
\varrho_{\mathsf{AA'}}^{(D,N)}&\nmsss=\nmsss&\frac{1}{\mathcal{N}_{D}^{(N)}}
\left[\sum_{i=0}^{N}\left(\mathcal{P}_{i}^{(N)}
+\mathcal{\overline{P}}^{(N)}_{i}\right)_{\mathsf{A}}\ot
\left(\mo{X_{D}^{(N)T_{i}}}^{T_{i}}\right)_{\mathsf{A}'}\right.\nonumber\\
&&\left.+\left(\ket{0}\!\bra{1}^{\ot N}+\ket{1}\!\bra{0}^{\ot
N}\right)_{\mathsf{A}}\ot
\left(X_{D}^{(N)}\right)_{\mathsf{A}'}\right],
\end{eqnarray}
where the subscripts $\mathsf{A}$ and $\mathsf{A}'$ are indicated
to distinguish their key and shield parts, respectively. However,
for the sake of clarity, in further considerations these
subscripts will be omitted.

The normalization factor $\mathcal{N}_{D}^{(N)}$ appearing in Eq. \eqref{Constr1}
is given by
\begin{equation}\label{normfactor}
\mathcal{N}_{D}^{(N)}=2\frac{(N+1)D^{N}+2D-4}{D^{N}+2D-4}.
\end{equation}

At the beginning we need to show that the matrices
$\varrho_{\mathsf{AA}'}^{(D,N)}$ really represent quantum states,
i.e., they are positive (the normalization condition is already
satisfied). Firstly, let us notice that the blocks corresponding
to $\mathcal{P}_{0}$ and $\overline{\mathcal{P}}_{0}$ and the two
off--diagonal blocks in Eq. \eqref{Constr1} constitute a matrix of
the form
$\mathcal{M}_{2}\big(\big|X_{D}^{(N)}\big|,X_{D}^{(N)}\big)$ (see
Lemma A.1 for the definition of $\mathcal{M}_{N}$), positivity of
which is guaranteed by Lemma A.1. Thus the only thing we need to
deal with is to show that the remaining blocks lying on the
diagonal of $\varrho_{\mathsf{AA}'}^{(D,N)}$ are positive. To
achieve this goal, below we prove a more general lemma.

{\it Lemma V.1.} Let $X_{D}^{(N)}$ be defined by Eq. (\ref{XDn}).
Then the matrices $\big|X_{D}^{(N)T_{k}}\big|^{T_{l}}$ are
positive semi--definite for all $k,l=1,\ldots,N$.

{\it Proof.} Noticing that $R_{D}^{(N)}$ is diagonal for arbitrary
$D$ and $N$, the partial transposition of $X_{D}^{(N)}$ with
respect to the $k$--th subsystem may be written as
$X_{D}^{(N)T_{k}}=[1/(D^{N}+2D-4)]\big(S^{(N)T_{k}}_{D}-R_{D}^{(N)}\big)$,
where $S_{D}^{(N)}$ is defined as
\begin{equation}\label{S}
S_{D}^{(N)}=\mathbbm{1}_{D^{N}}+DP_{D,N}^{(+)} -2R_{D}^{(N)}.
\end{equation}
As we will see below $S_{D}^{(N)T_{k}}$ is positive for any
$k=1,\ldots,N$ and $S_{D}^{(N)T_{k}}R_{D}^{(N)}=0$. Consequently,
the absolute value of $X_{D}^{(N)T_{k}}$ may be obtained by simple
changing the sign before the projector $R_{D}^{(N)}$. To prove
positivity of $S_{D}^{(N)T_{k}}$ let
$\ket{\psi}=\sum_{i_{1},\ldots,i_{N}}^{D-1}\alpha_{i_{1}\ldots
i_{N}}\ket{i_{1}\ldots i_{N}}$ denote an arbitrary vector from
$(\mathbb{C}^{D})^{\ot N}$ written in the standard basis of
$(\mathbb{C}^{D})^{\ot N}$. Then we have
\begin{eqnarray}\label{pos_of_S}
\Br{\psi}S_{D}^{(N)T_{k}}\Ke{\psi}
&\nmss=\nmss&\sum_{i\neq j}
\alpha_{i\ldots j\ldots i}^{*}\alpha_{j\ldots i\ldots
j}+\sum_{(i_{1},\ldots,i_{N})\in I}
\left|\alpha_{i_{1}\ldots
i_{N}}\right|^{2}\nonumber\\
&\nmss=\nmss&\hspace{-0.3cm}\sum_{(i_{1},\ldots,i_{N})\in
I_{k}}\hspace{-0.3cm}
\left|\alpha_{i_{1}\ldots i_{N}}\right|^{2} +
\frac{1}{2}\sum_{i\neq j}
\left|\alpha_{i\ldots j\ldots i}+ \alpha_{j\ldots i \ldots
j}\right|^{2}\nonumber\\
&\nmss\geq\nmss & 0.
\end{eqnarray}
Here the notation $\alpha_{i\ldots j\ldots i}$ means that all
indices of $\alpha$s excluding the $k$--th one ($k$ stands for the
number of subsystem being partially transposed) are equal.
Moreover, as previously $I$ denotes the set of all sequences
$(i_{1},\ldots,i_{N})$ except the cases when $i_{1}=\ldots=i_{N}$,
while $I_{k}$ is the set $I$ minus all sequences in which all
indices but the one on $k$--th position are equal.

As the value of $k$ is not specified, the above considerations
holds for any $k=1,\ldots,N$. Furthermore, using the same
reasoning one can also prove positiveness of $S_{D}^{(N)}$ being
transposed with respect to any subset of different subsystems
(besides the full transposition). This fact will be utilized below.

By virtue of the positiveness of $S_{D}^{(N)T_{k}}$ we have that
$\big|X_{D}^{(N)T_{k}}\big|=[1/(D^{N}+2D-4)]\big(S_{D}^{(N)T_{k}}+R_{D}^{(N)}\big)$
for any $k=1,\ldots,N$. Therefore the partial transposition of the latter with respect
to the $l$--th subsystem gives
\begin{equation}\label{cos}
\left|X_{D}^{(N)T_{k}}\right|^{T_{l}}=\frac{1}{D^{N}+2D-4}
\left(S_{D}^{(N)T_{k,l}}+R_{D}^{(N)}
\right),
\end{equation}
where $T_{k,l}$ denotes the partial transposition with respect to
two single subsystems $A_{k}'$ and $A_{l}'$.
Now we can distinguish two cases, namely, if $k=l$ and $k\neq l$.
In the first one, double partial transpositions with
respect to the same subsystem is just an identity map. Consequently from Eqs.
\eqref{projektory1}, \eqref{projektory2}, and \eqref{S}, one has
\begin{equation}\label{Constr5}
\left|X_{D}^{(N)T_{k}}\right|^{T_{k}}=
\frac{1}{D^{N}+2D-4}
\left(Q_{D}^{(N)}+DP^{(+)}_{D,N}\right),
\end{equation}
Now the right--hand side of Eq. (\ref{Constr5}) is a linear combination of
two positive operators and thus is positive.
We have still left the second case, that is, when $k\neq l$. To resolve it we may
use the remark made above, saying that
the partial transposition of $S_{D}^{(N)}$ with respect to arbitrary
non only one--partite subsystem is a positive matrix. This ends the proof. $\blacksquare$

Thus we have just proven that $\varrho_{\mathsf{AA}'}^{(D,N)}$ indeed represent
quantum states. Now, our aim is to prove that on the one hand they are bound entangled
and on the other hand they have nonzero distillable key. This purpose will be achieved
in two steps. Firstly we show that partial transposition with respect to any
elementary subsystem $(A_{i}A_{i}')$ of $\varrho_{\mathsf{AA}'}^{(D,N)}$ is positive.
Obviously, this does not confirm that the states are bound entangled since
we do not even know they are entangled. However, the latter may be proven
by showing that $K_{D}$ of these states is nonzero for $D\geq 3$.

Firstly, we concentrate on the positivity of all partial
transpositions of $\varrho_{\mathsf{A}\mathsf{A}'}^{(D,N)}$. To
gain a better look on the problem let us consider a particular
example of such a partial transposition, namely,
$\varrho_{\mathsf{A}\mathsf{A}'}^{(D,3)T_{3}}$. From Eq.
\eqref{Constr1} it follows that

\begin{widetext}

\begin{equation}\label{Ex_PT}
\varrho_{\mathsf{A}\mathsf{A}'}^{(D,3)T_{3}}=\frac{1}{\mathcal{N}_{D}^{(3)}}\left[
\begin{array}{cccccccc}
\left|X_{D}^{(3)}\right|^{T_{3}} & \hspace{-0.3cm}0 & \hspace{-0.5cm}0 & \hspace{-0.8cm}0 & \hspace{-0.8cm}0 & \hspace{-0.8cm}0 & \hspace{-0.5cm}0 & \hspace{-0.4cm}0 \\
0   & \hspace{-0.3cm}\left|X^{(3)T_{3}}_{D}\right|  & \hspace{-0.5cm}0 & \hspace{-0.8cm}0 & \hspace{-0.8cm}0 & \hspace{-0.8cm}0 & \hspace{-0.5cm}X^{(3)T_{3}}_{D} & \hspace{-0.4cm}0 \\
0   & \hspace{-0.3cm}0 & \hspace{-0.5cm}\left|X^{(3)T_{2}}_{D}\right|^{T_{2,3}} & \hspace{-0.8cm}0 & \hspace{-0.7cm}0 & \hspace{-0.7cm}0 & \hspace{-0.5cm}0 & \hspace{-0.4cm}0 \\
0   & \hspace{-0.3cm}0 & \hspace{-0.5cm}0 & \hspace{-0.8cm}\left|X^{(3)T_{1}}_{D}\right|^{T_{1,3}} & \hspace{-0.7cm}0 & \hspace{-0.7cm}0 & \hspace{-0.5cm}0 & \hspace{-0.4cm}0 \\
0   & \hspace{-0.3cm}0 & \hspace{-0.5cm}0 & \hspace{-0.8cm}0 & \hspace{-0.7cm}\left|X^{(3)T_{1}}_{D}\right|^{T_{1,3}} & \hspace{-0.7cm}0 & \hspace{-0.5cm}0 & \hspace{-0.4cm}0 \\
0   & \hspace{-0.3cm}0 & \hspace{-0.5cm}0 & \hspace{-0.8cm}0 & \hspace{-0.7cm}0 & \hspace{-0.7cm}\left|X^{(3)T_{2}}_{D}\right|^{T_{2,3}} & \hspace{-0.5cm}0 & \hspace{-0.4cm}0 \\
0   & \hspace{-0.3cm}X^{(3)T_{3}}_{D} & \hspace{-0.5cm}0 & \hspace{-0.8cm}0 & \hspace{-0.7cm}0 & \hspace{-0.7cm}0 & \hspace{-0.5cm}\left|X^{(3)T_{3}}_{D}\right| & \hspace{-0.4cm}0 \\
0   & \hspace{-0.3cm}0 & \hspace{-0.5cm}0 & \hspace{-0.8cm}0 & \hspace{-0.7cm}0 & \hspace{-0.7cm}0 & \hspace{-0.5cm}0 & \hspace{-0.4cm}\left|X_{D}^{(3)}\right|^{T_{3}}
\end{array}
\right].
\end{equation}

\end{widetext}

As, due to Lemma A.1, the square matrix consisting of two diagonal
and two off--diagonal blocks of
$\varrho_{\mathsf{AA}'}^{(D,N)T_{i}}$ (cf. Eq. \eqref{Ex_PT})
i.e., the matrix
$\mathcal{M}_{2}(\big|X^{(N)T_{i}}_{D}\big|,X^{(N)T_{i}}_{D} )$,
is already positive, what we need to prove is positivity
of $\big|X_{D}^{(N)}\big|^{T_{i}}$ and
$\big|X_{D}^{(N)T_{i}}\big|^{T_{j,k}}$ for any $i,j,k=1,\ldots,N$.
Let us therefore prove the following lemma.

{\it Lemma V.2.} Let $X_{D}^{(N)}$ be given by Eq. \eqref{XDn}.
Then for any $i,j,k=1,\ldots,N$ it holds that
\begin{equation}
\left|X_{D}^{(N)}\right|^{T_{i}}\geq 0,\qquad \left|X_{D}^{(N)T_{i}}\right|^{T_{j,k}}\geq 0.
\end{equation}

{\it Proof.} Due to the definition of $X_{D}^{(N)}$ its absolute
value may be calculated simply by changing a sign before
$P_{D}^{(N)}$, giving
\begin{equation}
\left|X_{D}^{(N)}\right|=\frac{1}{D^{N}+2D-4}\left[(D-2)P^{(+)}_{D,N}+2P_{D}^{(N)}+Q_{D}^{(N)}\right].
\end{equation}
Application of partial transposition with respect to the $i$th
subsystem followed by substitution of Eq. \eqref{projektory1}
leads us to
\begin{equation}
\left|X_{D}^{(N)}\right|^{T_{i}}=\frac{1}{D^{N}+2D-4}\left[\mathbbm{1}_{D}+R_{D}^{(N)}+(D-4)P_{D,N}^{(+)T_{i}}
\right]
\end{equation}
for any $i=1,\ldots,N$. One may easily check that eigenvalues of
$P_{D,N}^{(+)T_{i}}$ belong to the interval $[-1/D,1/D]$ and
therefore the matrix $\mathbbm{1}_{D^{N}}+(D-4)P_{D,N}^{(+)T_{i}}$
is always positive. This, together with the fact that
$R_{D}^{(N)}\geq 0$, implies positivity of
$\big|X_{D}^{(N)}\big|^{T_{i}}$ for any $i=1,\ldots,N$.

The second fact of the lemma may be proven just by noting that by
virtue of Eq. (\ref{cos}) it holds
\begin{equation}
\left|X_{D}^{(N)T_{i}}\right|^{T_{j,k}}=\frac{1}{D^{N}+2D-4}
\left(S_{D}^{(N)T_{i,j,k}}+R_{D}^{(N)}
\right).
\end{equation}
As stated previously in the proof of Lemma V.1, the partial transposition
of $S_{D}^{(N)}$ with respect to arbitrary subsystems is positive.
This concludes the proof. $\blacksquare$

The above lemma proves actually that all the partial
transpositions $\varrho_{\mathsf{AA}'}^{(D,N)T_{i}}$
$(i=1,\ldots,N)$ are positive. Therefore, the states
$\varrho_{\mathsf{AA}'}^{(D,N)}$ are bound entangled, of course
provided that they are entangled. This is because, due to the
result of Ref. \cite{bound}, positive partial transpositions with
respect to any elementary subsystem makes it impossible to distill
$k$--partite ($k=2,\ldots,N$) GHZ entanglement among any group of
parties.





Let us now pass to the proof that any state $\varrho_{\mathsf{AA}'}^{(D,N)}$
for $D\geq 3$ has nonzero $K_{D}$. For this purpose we show that using
the protocol from Section \ref{LOCCProtocol} we can produce a state that is closer
to some multipartite private state out of copies of $\varrho_{\mathsf{AA}'}^{(D,N)}$.
As we will show below we need to use as many copies as it is necessary to
make the quantity appearing on the right--hand side of Eq. \eqref{DWbound3}
strictly positive.

Application of the recursive LOCC protocol presented in Section
\ref{LOCCProtocol} to $k$ copies of
$\varrho_{\mathsf{AA}'}^{(D,N)}$ gives with probability
$p_{D,N}^{(k)}=2^{k-1}\mathcal{N}_{D,N}^{(k)}/\big(\mathcal{N}_{D}^{(N)}\big)^{k}$
the following state
\begin{eqnarray}\label{Theta1}
\Theta_{\mathsf{AA'}}^{(N,k)}&\nmss=\nmss&\frac{1}{\mathcal{N}_{D,N}^{(k)}}
\left[
\sum_{i=0}^{N}\left(\mathcal{P}_{i}^{(N)}+\mathcal{\overline{P}}^{(N)}_{i}\right)\ot
\left(\mo{X_{D}^{(N)T_{i}}}^{T_{i}}\right)^{\ot k}\right.\nonumber\\
&&\left.+\left(\ket{0}\!\bra{1}^{\ot N}+\ket{1}\!\bra{0}^{\ot N
}\right)\ot \left(X_{D}^{(N)}\right)^{\ot k}\right],
\end{eqnarray}
where $\mathcal{N}_{D,N}^{(k)}$ is a normalization factor given by
\begin{eqnarray}\label{normalization}
\mathcal{N}_{D,N}^{(k)}=
2\left[1+N\left(\frac{D^{N}}{D^{N}+2D-4}\right)^{k}\right].
\end{eqnarray}
Now, to simplify the considerations we can utilize the privacy
squeezing (see Section \ref{PrivacySqueezing}) to the obtained
states $\Theta_{\mathsf{AA'}}^{(N,k)}$. Namely, due to Lemma III.1
there exist such twistings $U_{t}^{(k)}$ that after application to
$\Theta_{\mathsf{AA'}}^{(N,k)}$ and tracing out the $\mathsf{A}'$
subsystem one arrives at the following class of $N$--qubit states
\begin{eqnarray}\label{Question}
\widetilde{\Theta}_{\mathsf{A}}^{(N,k)}&\nmss=\nmss&\frac{1}{\mathcal{N}_{D,N}^{(k)}}\left[
\sum_{i=0}^{N}\left(\mathcal{P}_{i}^{(N)}+\mathcal{\overline{P}}^{(N)}_{i}\right)
\norsl{\mo{X_{D}^{(N)T_{i}}}^{T_{i}}}^{k}\right.\nonumber\\
&&\left.+\left(\ket{0}\!\bra{1}^{\ot N}+\ket{1}\!\bra{0}^{\ot N }\right)
\norsl{X_{D}^{(N)}}^{k}\right].
\end{eqnarray}
In other words, after 'rotation' with $U_{t}^{(k)}$ and throwing
out the $\mathsf{A}'$ subsystem we get a so--called privacy squeezed
state, i.e., the one in which blocks are replaced with their
norms. We also know from Theorem IV.4 that the distillable key of
the \textsf{c}q state obtained from the privacy squeezed state
$\widetilde{\Theta}_{\mathsf{A}}^{(N,k)}$ (measurement is
performed in the same basis as twisting) cannot be greater than
the distillable key of $\Theta_{\mathsf{AA'}}^{(N,k)}$.

From Eq. \eqref{normalization} it follows that since $D^{N}+2D-4>
D^{N}$ for any $D\geq 3$ one has $\mathcal{N}_{D,N}^{(k)}\to 2$,
while for $D=2$, $\mathcal{N}_{2,N}^{(k)}\to 2(N+1)$. In turn this
means that for the off--diagonal elements of
$\widetilde{\Theta}_{\mathsf{A}}^{(N,k)}$ one has that
\begin{equation}
\frac{1}{\mathcal{N}_{D,N}^{(k)}}\left\|X_{D}^{(N)}\right\|_{1}^{k}=\frac{1}{\mathcal{N}_{D,N}^{(k)}}
\stackrel{k\to\infty}{\LRA} \frac{1}{2}
\end{equation}
%
with $D\geq 3$. By virtue of Theorem III.3 one infers that the
more copies of $\varrho_{\mathsf{AA'}}^{(D,N)}$ we put into the
recurrence protocol, the closer we are to some multipartite
private state. This also means that with $k\to \infty$ the
sequence of states $\widetilde{\Theta}_{\mathsf{A}}^{(N,k)}$ goes
to $GHZ$ state $P_{2,N}^{(+)}$, however, for $D\geq 3$.

Now, to bound from below the distillable key of
$\Theta_{\mathsf{AA'}}^{(N,k)}$ according to the prescription
given above we need to calculate a \textsf{c}q state of the
privacy squeezed state $\widetilde{\Theta}_{\mathsf{A}}^{(N,k)}$.
(The \textsf{c}q state is found here with respect to the basis in
which the original state is defined.) Simple algebra gives
\begin{eqnarray}\label{Constr1CCQWielocz}
\widetilde{\Theta}_{\mathsf{A}E}^{(\mathsf{c}\mathrm{q})}&\nmss=\nmss&\frac{1}{\mathcal{N}_{D,N}^{(k)}}
\left[R_{2}^{(N)}
\ot\proj{E_{0}}+\left(\frac{D^{N}}{D^{N}+2D-4}\right)^{k}\right.\nonumber\\
&&\left.\times\sum_{j=1}^{N}\left(P_{j}^{(N)}\ot\proj{E_{j}}
+\overline{P}_{j}^{(N)}\ot\proj{\overline{E}_{j}}\right)\right],\nonumber\\
\end{eqnarray}
where $\ket{E_{0}}$, $\ket{E_{j}}$, and $\ket{\overline{E}_{j}}$
constitute a set of orthonormal states held by Eve. One notices
immediately that
$\widetilde{\Theta}_{\mathsf{A}E}^{(\mathsf{c}\mathrm{q})}$ tends
to the multipartite ideal \textsf{c}q state (see Eq.
\eqref{idealcq}) for any integer $D\geq 3$ whenever $k\to \infty$.

\begin{figure}[h!]
(a)\includegraphics[width=6cm]{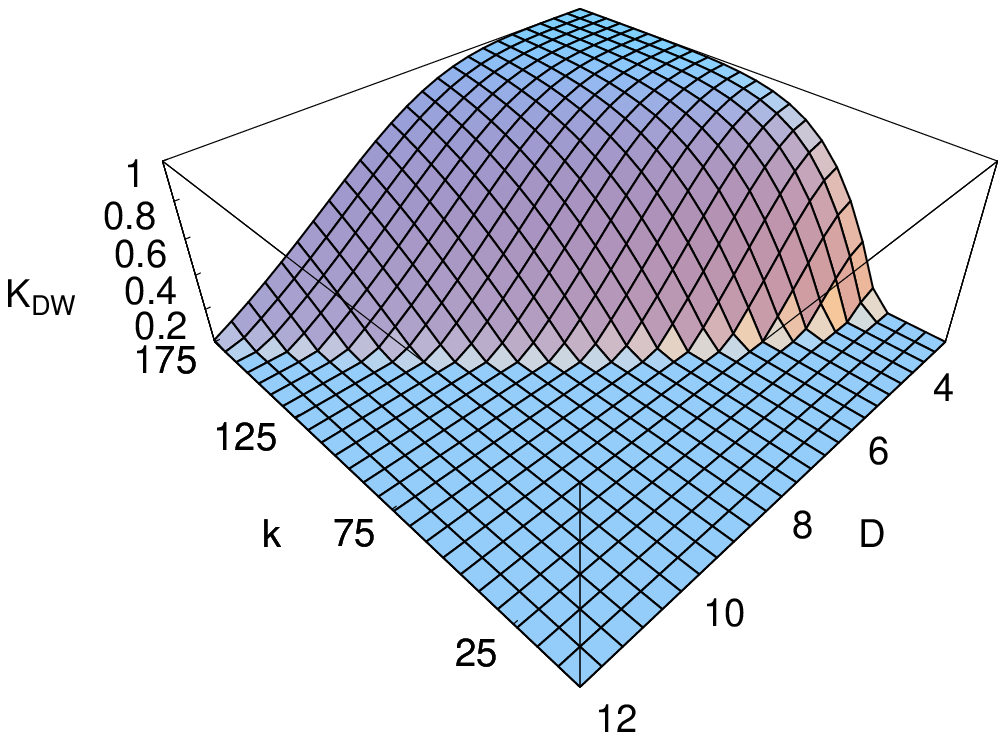}\\
(b)\includegraphics[width=6cm]{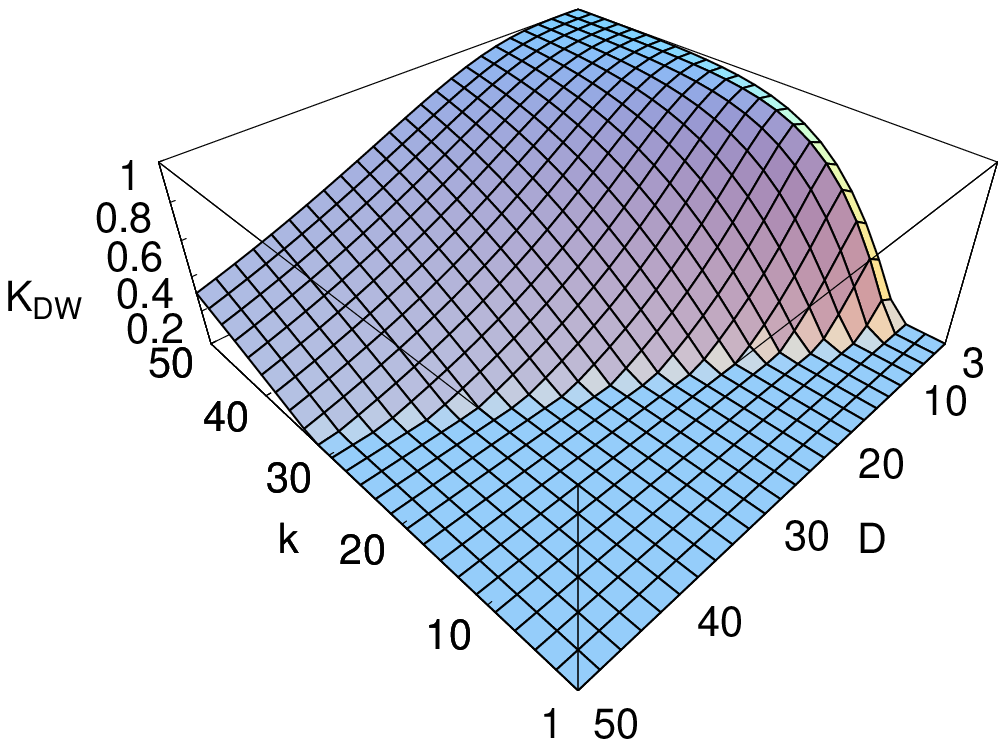} \caption{An
exemplary plot of $K_{DW}\equiv
I(A_{1}\!:\!A_{2})(\Theta_{A_{1}A_{2}E}^{(\mathrm{ccq})})-I(A_{1}\!:\!E)(\Theta_{A_{1}A_{2}E}^{(\mathrm{ccq})})$
with $N=3$ (a) and for comparison in the case of $N=2$ (b), which
was discussed in Ref. \cite{PHRA}. For the sake of clarity,  zero
is put whenever the plotted function is less than zero. Notice
also that even though $k$ and $D$ are discrete parameters, the
graph is made as if $K_{DW}$ were a function of continuous
parameters. It follows from both the figures that the number of
parties $N$ significantly influences the obtained lower bound.
Namely, for $N=3$ one needs to spend more copies of a given state
to get nonzero values of $K_{DW}$.} \label{rys1Constr2}
\end{figure}

To find a lower bound on distillable key of
$\widetilde{\Theta}_{\mathsf{A}}^{(N,k)}$ we utilize Eq. \eqref{DWbound2}.
However, according to Eq. \eqref{DWbound2} one needs to calculate
the quantities $I(A_{i}\!:\!A_{j})$ for $i\neq j$ and
$I(A_{i}\!:\! E)$ for the respective reductions of
$\Theta_{\mathsf{A}E}^{(\mathsf{c}\mathrm{q})}$. Fortunately, the
initial states $\varrho_{\mathsf{AA}'}^{(D,N)}$ have such
symmetrical structure, preserved by the recurrence protocol and
the privacy squeezing, that makes all the quantities
$I(A_{i}\!:\!A_{j})$ $(i\neq j)$ equal (the same holds for
$I(A_{i}\!:\! E)$). Consequently, in view of the above, using Eq.
\eqref{DWbound2} and Theorem IV.4 (see Eq. \eqref{theoremIV4}), we
infer the following inequality
\begin{equation}\label{Constr2LowerBound}
K_{D}(\Theta_{\mathsf{AA}'}^{(N,k)})\geq
I(A_{1}\!:\!A_{2})(\Theta_{A_{1}A_{2}E}^{(\mathrm{ccq})})-I(A_{1}\!:\!E)(\Theta_{A_{1}A_{2}E}^{(\mathrm{ccq})}).
\end{equation}
irrespectively of number of parties $N$. Exemplary behaviour of the right--hand side
of Eq. \eqref{Constr2LowerBound} (denoted by $K_{DW}$)
in the function of $k$ and $D$ for $N=3$ is shown in Fig. \ref{rys1Constr2}a.
%
%
It is clear from Figure \ref{rys1Constr2}a that it is possible to
distill one secure bit of key from bound entangled states
$\Theta_{\mathsf{AA}'}^{(N,k)}$ for sufficiently large $k$. For
comparison, Fig. \ref{rys1Constr2}b contains a lower bound of the
distillable key in the case of $N=2$ discussed in Ref.
\cite{PHRANATO}.

\begin{figure}[h!]
(a)\includegraphics[width=7cm]{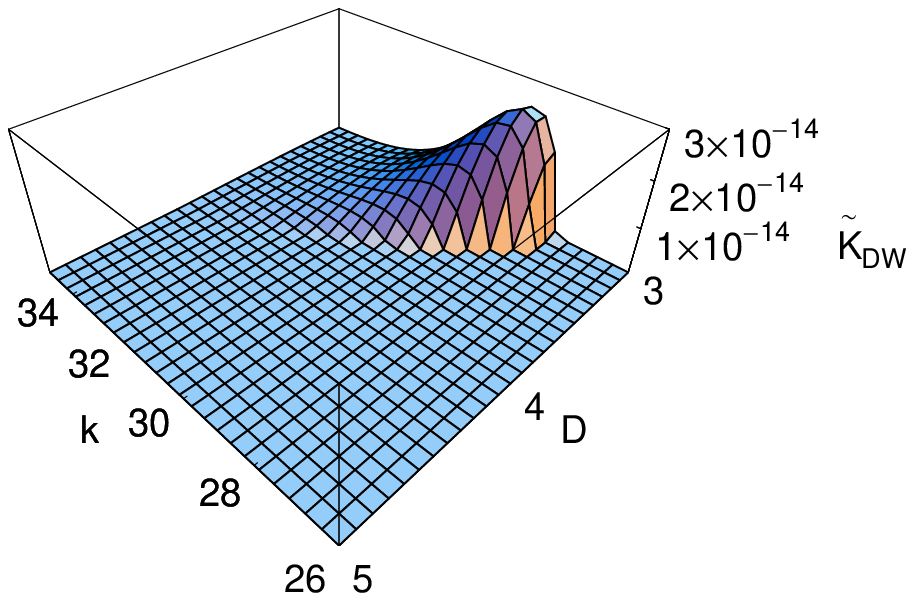}\\
(b)\includegraphics[width=6cm]{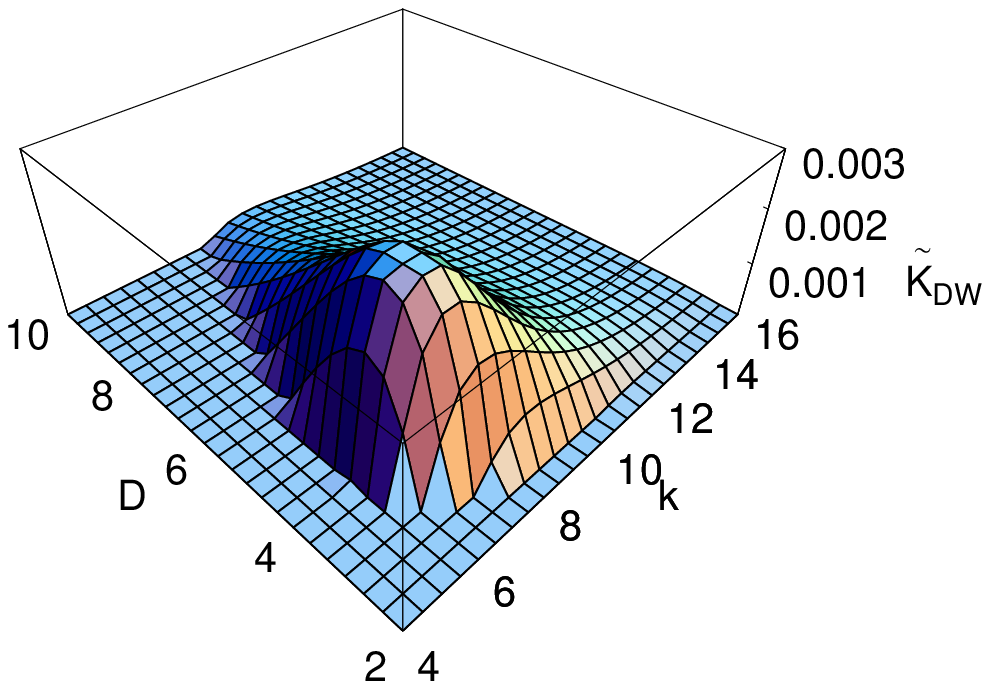} \caption{An
exemplary plot of $\widetilde{K}_{DW}\equiv p_{D,3}^{(k)}
\left[I(A_{1}\!:\!A_{2})(\Theta_{A_{1}A_{2}E}^{(\mathrm{ccq})})-
I(A_{1}\!:\!E)(\Theta_{A_{1}A_{2}E}^{(\mathrm{ccq})})\right]$ with
$N=3$. For comparison it is also presented the case of $N=2$ (b).
For the sake of clarity, zero is put whenever the plotted function
is less or equal to zero. Also, though both the parameters $k$ and
$D$ are integer, for convenience, the function
$\widetilde{K}_{DW}$ is plotted as if it were a function of
continuous $k$ and $D$. It is clear that for $N=3$ the lower bound
on distillable key is considerably smaller. } \label{rys2Constr2}
\end{figure}

We can also investigate the lower bound on $K_{D}$
for the initial states $\varrho_{\mathsf{AA}'}^{(D,N)}$. However, in
this case we need to take into account the probability $p_{D,N}^{(k)}$.
In this way we arrive at
\begin{eqnarray}\label{Constr2LowerBound2}
K_{D}(\varrho_{\mathsf{AA}'}^{(D,N)})&\nmss\geq\nmss& p_{D,N}^{(k)}\left[ I(A_{1}\!:\!A_{2})(\Theta_{A_{1}A_{2}E}^{(\mathrm{ccq})})\right.\nonumber\\
&&\left.\hspace{1cm}-I(A_{1}\!:\!E)(\Theta_{A_{1}A_{2}E}^{(\mathrm{ccq})})\right].
\end{eqnarray}
Figure \ref{rys2Constr2}a presents exemplary behaviour
of the function appearing on the right--hand side
of Eq. \eqref{Constr2LowerBound2} (denoted by $\widetilde{K}_{DW}$) for $N=3$.
For comparison, in Figure \ref{rys2Constr2}b it is also plotted
the same function in the case of $N=2$ (this case was discussed in Ref. \cite{PHRA}).

%
%
Let us conclude the first construction with discussion of some of its general
properties. Above we used a particular class of matrices $X_{D}^{(N)}$
(defined in Eq. (75)), however, it seems interesting to ask wether there are
other matrices than $X_{D}^{(N)}$ that could be used in the construction.
In what follows we provide some constraints that the general matrix,
hereafter denoted by $Z_{D}^{(N)}$, must obey to be useful for
purposes of the construction. The first important condition is that
the trace norm of $Z_{D}^{(N)}$ has to be strictly larger than the trace norm of
$\big|Z_{D}^{(N)T_{i}}\big|^{T_{i}}$ for all $i=1,...,N$. This guarantees
convergence (in the trace norm) of the output states of the recursive
LOCC protocol (given in Sec. IV.C) to some multipartite
private states. Another crucial conditions
are $\big|Z_{D}^{(N) T_{i}}\big|^{T_{i}}\geq 0$ and
$\big|Z_{D}^{(N)}\big|^{T_{i}}\geq 0$ for all $i=1,\ldots,N$.
The first one is necessary for $\varrho_{\mathsf{AA}'}^{(D,N)}$
(when constructed with the matrix $Z_{D}^{(N)}$) to be positive,
while the second one allows to prove that $\varrho_{\mathsf{AA}'}^{(D,N)}$
have positive partial transposition with respect to any elementary subsystem.\\

{\it Lemma V.3.} Assume that $Z_{D}^{(N)}$ is arbitrary matrix
acting on $(\mathbb{C}^{D})^{\ot N}$ and that the following conditions
\begin{enumerate}
  \item[(i)] $\norsl{Z_{D}^{(N)}}> \norsl{\left|Z_{D}^{(N)T_{i}}\right|^{T_{i}}}$ for all $i=1,\ldots,N$,
  \item[(ii)] $\left|Z_{D}^{(N)T_{i}}\right|^{T_{i}}\geq 0$ for all $i=1,\ldots,N$,
  \item[(iii)] $\left|Z_{D}^{(N)}\right|^{T_{i}}\geq 0$ for all $i=1,\ldots,N$
\end{enumerate}
{are satisfied. Then $Z_{D}^{(N)}\ngeq 0$ and $Z_{D}^{(N)T_{i}}\ngeq 0$ for all $i=1,\ldots,N$.}\\

{\it Proof.} ({\it ad absurdum}) We divide the proof into three parts:
\begin{enumerate}
  \item[(i)] Assume that $Z_{D}^{(N)}\geq 0$ and $Z_{D}^{(N)T_{i}}\ngeq 0$ for any $i=1,\ldots,N$.
  Then one can see that $\big|Z_{D}^{(N)}\big|^{T_{i}}=Z_{D}^{(N)T_{i}}\ngeq 0$ for any choice of $i$.
  However, this contradicts the third assumption.
  \item[(ii)] Assume that $Z_{D}^{(N)}\ngeq 0$ and there exists such $k$ that $Z_{D}^{(N)T_{k}}\geq 0$. Now,
  one obtains $\big|Z_{D}^{(N)T_{k}}\big|^{T_{k}}=Z_{D}^{(N)}\ngeq 0$. Of course, this is in contradiction to the second assumption.
  \item[(iii)] Finally, assume that $Z_{D}^{(N)}\geq 0$ and that there exists such $k$ that
  $Z_{D}^{(N)T_{k}}\geq 0$. Then $\norsl{\big|Z_{D}^{(N)T_{k}}\big|^{T_{k}}}=\norsl{Z_{D}^{(N)}}$. This contradicts the first assumption.
      $\blacksquare$
\end{enumerate}
This lemma says that a matrix can be used in the above construction
if it is not positive and all its elementary partial transpositions are not positive.
Thus, in particular, a general density matrix is not suitable for this construction.

\subsection{The second construction} \label{Giechazety2}
The crucial ideas behind the second construction are actually the same as in the
case of the first one, however, considerations will be a little bit more sophisticated.

Let us first define the analog of $X_{D}^{(N)}$ from the first construction
to be
\begin{equation}\label{tildeX}
\widetilde{X}^{(N)}_{D}=\sum_{i,j=0}^{D-1}u_{ij}\ket{i}\!\bra{j}^{\ot
N},
\end{equation}
where we assume that $u_{ij}$ are elements of some $D\times D$
general unitary or unitary Hermitian matrix, hereafter denoted by $U_{D}$.
Thus $\widetilde{X}_{D}^{(N)}$ is an embedding of $U_{D}\in
M_{D}(\mathbb{C})$ ($M_{D}(\mathbb{C})$ denotes the set of
$D\times D$ matrices with complex entries) in
$M_{D^{N}}(\mathbb{C})$ and therefore
\begin{equation}\label{relacion}
\left|\widetilde{X}_{D}^{(N)}\right|=R_{D}^{(N)}.
\end{equation}
For further simplicity we also impose the condition that
$|u_{ij}|=1/\sqrt{D}$ for $i,j=0,\ldots,D-1$, however whenever
possible all proofs will be given assuming that $U_{D}$ is
a general unitary matrix.

It should be also pointed out that the distinction on unitary or
unitary and Hermitian matrices $U_{D}$ made above plays an important
role here. This comes from the LOCC protocol presented in Section
\ref{LOCCProtocol} as in the case of unitary but not Hermitian
matrices it needs to be slightly modified. Namely, in its last
step all the parties keep the state only if all zeros occurred.

A particular example of a unitary but in general not Hermitian
matrix satisfying the above condition is the matrix
$\widetilde{V}_{D}=(1/\sqrt{D})V_{D}$, where $V_{D}$ denotes the
Vandermonde matrix of solutions to
the equation $z^{D}-1=0$ with $z\in\mathbb{C}$. As one knows the
solutions are of the form $\omega_{k}=\mathrm{e}^{2\pi
\mathrm{i}k/D}$ $(k=0,\ldots,D-1)$. It is then clear that
$\widetilde{V}_{D}$ is a unitary matrix for any $D\geq 2$, however
not always a Hermitian one. For instance, in the particular case
of $D=2$ one easily recognizes that $\widetilde{V}_{2}$ is the
known Hadamard matrix. A good example of some unitary and
Hermitian matrix is $k$th tensor power of $\widetilde{V}_{2}$.
Since $\widetilde{V}_{2}$ is unitary and Hermitian any matrix
of the form $\widetilde{V}_{2}^{\ot k}$ is also unitary and
Hermitian.

Now, let us consider following family of matrices
\begin{eqnarray}\label{SecondConstr}
\widetilde{\varrho}^{(D,N)}_{\mathsf{A}\mathsf{A'}}&\nmss=\nmss&\frac{1}{\widetilde{\mathcal{N}}^{(N)}_{D}}
\left[\sum_{j=0}^{N}\left(\mathcal{P}_{j}^{(N)}
+\overline{\mathcal{P}}_{j}^{(N)}\right)\ot\sum_{i=1}^{N}\left|\widetilde{X}_{D,i}^{(N)T_{j}}\right|\right.\nonumber\\
&&\left.+\ket{ 0}\!\bra{1}^{\ot
N}\ot\sum_{i=1}^{N}\widetilde{X}_{D,i}^{(N)}+\ket{
1}\!\bra{0}^{\ot N}\ot\sum_{i=1}^{N}\widetilde{X}_{D,i}^{(N)\dagger}\right],\nonumber\\
\end{eqnarray}
where $\widetilde{\mathcal{N}}^{(N)}_{D}$ stands for the
normalization factor, which for arbitrary unitary $U_{D}$ is given
by
\begin{eqnarray}
\widetilde{\mathcal{N}}_{D}^{(N)}=2N\left(D+N\sum_{i,j=0}^{D-1}|u_{ij}|\right).
\end{eqnarray}
Obviously for $\widetilde{X}_{D}^{(N)}$ that comes from unitary
Hermitian $U_{D}$ the conjugation in the last term in Eq.
\eqref{SecondConstr} may be omitted. Moreover, taking into account
the assumption that $|u_{ij}|=1/\sqrt{D}$, the normalization
factor becomes
$\widetilde{\mathcal{N}}_{D}^{(N)}=2ND(1+N\sqrt{D})$.

As in the case of the first construction, we need to prove that
$\widetilde{\varrho}^{(D,N)}_{\mathsf{A}\mathsf{A'}}$ represent
quantum states. Moreover, we show also that they have positive
partial transpositions with respect to any elementary subsystem.
From Eq. \eqref{SecondConstr} it follows that to prove positivity
of $\widetilde{\varrho}_{\mathsf{AA}'}^{(D,N)}$ one has to show
that the inequalities
\begin{equation}\label{Nier1}
\left|\sum_{i=1}^{N}\widetilde{X}_{D,i}^{(N)}\right|\leq
\sum_{i=1}^{N}\left|\widetilde{X}_{D,i}^{(N)}\right|
\end{equation}
are satisfied. Then simply utilizing Lemma A.1 and noting that the
remaining blocks lying on the diagonal of
$\widetilde{\varrho}_{\mathsf{AA}'}^{(D,N)}$ are positive by
definition, the positivity of
$\widetilde{\varrho}_{\mathsf{AA}'}^{(D,N)}$ is proved.

To deal with the problem of positivity of partial transpositions
let us look on the particular example of form of
$\widetilde{\varrho}^{(D,3)T_{3}}_{\mathsf{A}\mathsf{A'}}$. From
Eq. \eqref{SecondConstr} one infers that
\begin{widetext}

\begin{equation}\label{EX_PT2}
\widetilde{\varrho}^{(D,3)T_{3}}_{\mathsf{A}\mathsf{A'}}=\frac{1}{\widetilde{\mathcal{N}}^{(3)}_{D}}\left[
\begin{array}{cccccccc}
\displaystyle\sum_{i=1}^{3}\Big|\widetilde{X}_{D,i}^{(3)}\Big| &
\hspace{-0.4cm}0 & \hspace{-0.4cm}0 & \hspace{-0.4cm}0
&\hspace{-0.4cm} 0 &
\hspace{-0.4cm}0 &\hspace{-0.4cm} 0 & \hspace{-0.4cm}0\\

0 &
\hspace{-0.4cm}\displaystyle\sum_{i=1}^{3}\Big|\widetilde{X}_{D,i}^{(3)T_{3}}\Big|
& \hspace{-0.4cm}0 & \hspace{-0.4cm}0 & \hspace{-0.4cm}0 &
\hspace{-0.4cm}0 &
\hspace{-0.4cm}\displaystyle\sum_{i=1}^{3}\widetilde{X}_{D,i}^{(3)T_{3}} &\hspace{-0.4cm} 0\\

0 & \hspace{-0.4cm}0 &
\hspace{-0.4cm}\displaystyle\sum_{i=1}^{3}\Big|\widetilde{X}_{D,i}^{(3)T_{2}}\Big|
& \hspace{-0.4cm}0
& \hspace{-0.4cm}0 & \hspace{-0.4cm}0 & \hspace{-0.4cm}0 & \hspace{-0.4cm}0 \\

0 & \hspace{-0.4cm}0 & \hspace{-0.4cm}0 &
\hspace{-0.4cm}\displaystyle\sum_{i=1}^{3}\Big|\widetilde{X}_{D,i}^{(3)T_{1}}\Big|
& \hspace{-0.4cm}0
& \hspace{-0.4cm}0 & \hspace{-0.4cm}0 & \hspace{-0.4cm}0 \\

0 & \hspace{-0.4cm}0 &\hspace{-0.4cm} 0 & \hspace{-0.4cm}0 &
\hspace{-0.4cm}\displaystyle\sum_{i=1}^{3}\Big|\widetilde{X}_{D,i}^{(3)T_{1}}\Big|
& \hspace{-0.4cm}0 & \hspace{-0.4cm}0 & \hspace{-0.4cm}0\\

0 & \hspace{-0.4cm}0 & \hspace{-0.4cm}0 & \hspace{-0.4cm}0 &
\hspace{-0.4cm}0 &
\hspace{-0.4cm}\displaystyle\sum_{i=1}^{3}\Big|\widetilde{X}_{D,i}^{(3)T_{2}}\Big|
& \hspace{-0.4cm}0 & \hspace{-0.4cm}0\\

0 &
\hspace{-0.4cm}\displaystyle\sum_{i=1}^{3}\widetilde{X}_{D,i}^{(3)T_{3}\dagger}
& \hspace{-0.4cm}0 & \hspace{-0.4cm}0 & \hspace{-0.4cm}0 &
\hspace{-0.4cm}0 &
\hspace{-0.4cm}\displaystyle\sum_{i=1}^{3}\Big|\widetilde{X}_{D,i}^{(3)T_{3}}\Big|
& \hspace{-0.4cm}0 \\

0 & \hspace{-0.4cm}0 & \hspace{-0.4cm}0 & \hspace{-0.4cm}0 &
\hspace{-0.4cm}0 & \hspace{-0.4cm}0 & \hspace{-0.4cm}0 &
\hspace{-0.4cm}\displaystyle\sum_{i=1}^{3}\Big|\widetilde{X}_{D,i}^{(3)}\Big|
\end{array}
\right],
\end{equation}
\end{widetext}
where we used the fact that
$\big|\widetilde{X}_{D,i}^{(n)T_{j}}\big|$ are diagonal in the
standard basis and therefore are not affected by partial
transposition with respect to any subsystems.

To show positivity of $\widetilde{\varrho}_{\mathsf{AA}'}^{(D,N)}$
as well as its partial transpositions we prove the following
lemma.

{\it Lemma V.4.} Let $\widetilde{X}_{D}^{(N)}$ be defined as in
Eq. \eqref{tildeX}. Then the following equalities holds
\begin{equation}\label{Nier1}
\left|\sum_{i=1}^{N}\widetilde{X}_{D,i}^{(N)T_{j}}\right|=
\sum_{i=1}^{N}\left|\widetilde{X}_{D,i}^{(N)T_{j}}\right| \qquad
(j=0,\ldots,N).
\end{equation}
%
%
%
%
%
%
%
{\it Proof.} Firstly we start by the above statement for $j=0$.
For this purpose let us notice that its
left--hand side may be written as
\begin{equation}
\left|\sum_{i=1}^{N}\widetilde{X}_{D,i}^{(N)}\right|=
\left|N\sum_{k=0}^{D-1}u_{kk}\proj{k}^{\ot
N}+\sum_{i=1}^{N}\sum_{\substack{k,l=0\\k\neq
l}}^{D-1}u_{kl}(\ket{k}\!\bra{l}^{\ot N})^{T_{i}}\right|.
\end{equation}
Straightforward algebra shows that both terms under the sign of
absolute value are defined on orthogonal supports. Moreover,
all the partial transpositions in the second term are defined on
orthogonal supports. Both these facts allow us to write
\begin{equation}
\left|\sum_{i=1}^{N}\widetilde{X}_{D,i}^{(N)}\right|=N\sum_{k,l=0}^{D-1}\left|u_{kl}\right|\proj{l}^{\ot(i-1)}\ot
\proj{k}\ot\proj{l}^{\ot(N-i-1)}.
\end{equation}
One finds immediately that this equals the right--hand side of
(\ref{Nier1}), finishing the first part of the proof.

To show Eq. \eqref{Nier1} for $j=1,\ldots,N$ we need to perform a
little bit more sophisticated analysis. With the same reasoning as
in the case of the first inequality we can reduce the claimed
inequalities to the following
\begin{eqnarray}\label{a4}
&&\left|\widetilde{X}^{(N)}_{D}+(N-1)\sum_{k=0}^{D-1}u_{kk}\proj{k}^{\ot N}\right|\leq R_{D}^{(N)}
\nonumber\\
&&\hspace{1cm}+(N-1)\sum_{k=0}^{D-1}|u_{kk}|\proj{k}^{\ot N},
\end{eqnarray}
where we utilized Eq. \eqref{relacion}. One notices that the above
inequality may be further reduced to
\begin{equation}
\left|U_{D}+(N-1)\mathcal{D}\right|\leq\mathbbm{1}_{D}+(N-1)\left|\mathcal{D}\right|,
\end{equation}
where $\mathcal{D}$ denotes a diagonal matrix containing the
diagonal elements of $U_{D}$. Utilizing the fact that
$|u_{ij}|=1/\sqrt{D}$ for any $i,j=0,\ldots,D-1$, we infer that
$|\mathcal{D}|=(1/\sqrt{D})\mathbbm{1}_{D}$ and therefore
\begin{equation}\label{inequality}
\left|U+(N-1)\mathcal{D}\right|\leq
[1+(N-1)/\sqrt{D}]\mathbbm{1}_{D}.
\end{equation}
To prove this inequality we can utilize the polar decomposition to
its left--hand side. More precisely we can write
$\left|U+(N-1)\mathcal{D}\right|=V^{\dagger}U+(N-1)V^{\dagger}\mathcal{D}$
with $V$ denoting some unitary matrix. This allows us to write
\begin{eqnarray}\label{a5}
\bra{\Psi}\left|U+(N-1)\mathcal{D}\right|\ket{\Psi}&\nmss=\nmss&
\big|\bra{\Psi}\left|U+(N-1)\mathcal{D}\right|\ket{\Psi}\big|\nonumber\\
&\nmss\leq\nmss&\left|\bra{\Psi}V^{\dagger}U\ket{\Psi}\right|\nonumber\\
&&+(N-1)\left|\bra{\Psi}V^{\dagger}\mathcal{D}\ket{\Psi}\right|\nonumber\\
&\nmss\leq \nmss&1+(N-1)\left|\bra{\Psi}V^{\dagger}|\mathcal{D}|W\ket{\Psi}\right|\nonumber\\
&\nmss\leq \nmss&1+\frac{N-1}{\sqrt{D}},
\end{eqnarray}
where $\ket{\Psi}$ is an arbitrary normalized vector from
$\mathbb{C}^{D}$. The second and third inequality are consequences
of the fact that the product of unitary matrices is a unitary matrix
and that for any normalized $\ket{\psi}$ and unitary $U$ it holds
that $|\bra{\psi}U\ket{\psi}|\leq 1$. Moreover, we put here the
polar decomposition of $\mathcal{D}$, i.e.,
$\mathcal{D}=|\mathcal{D}|W$ with some unitary $W$. The last
inequality is also a result of application of aforementioned fact
that $|\mathcal{D}|=(1/\sqrt{D})\mathbbm{1}_{D}$.

Now, to finish the proof, it suffices to mention that the
resulting inequality is equivalent to \eqref{inequality}.
$\blacksquare$

From the above lemma it clearly follows that $\widetilde{\varrho}_{\mathsf{AA}'}^{(D,N)}$
represent quantum states for any $D\geq 2$ and $N\geq 2$, and they have positive partial transpositions
with respect to all elementary subsystems. The last thing we need to prove
is that the distillable key of $\widetilde{\varrho}_{\mathsf{AA}'}^{(D,N)}$
is nonzero. This would also imply that $\widetilde{\varrho}_{\mathsf{AA}'}^{(D,N)}$
represent entangled states.

Let us then apply the recursive protocol described previously in Section \ref{LOCCProtocol}
to $k$ copies of $\widetilde{\varrho}_{\mathsf{AA}'}^{(D,N)}$, obtaining
\begin{eqnarray}\label{SecondConstrLOCC}
\widetilde{\Theta}_{\mathsf{AA'}}^{(N,k)}&\nmss=\nmss&\frac{1}{\widetilde{\mathcal{N}}^{(k)}_{D,N}}
\left[\sum_{j=0}^{N}
\left(\mathcal{P}_{j}^{(N)}+\overline{\mathcal{P}}_{j}^{(N)}\right)\ot
\left(\sum_{i=1}^{N}\left|\widetilde{X}^{(N)T_{j}}_{D,i}\right|\right)^{\ot k}\right.\nonumber\\
&&+\ket{ 0}\!\bra{1}^{\ot N}\ot\left(\sum_{i=1}^{N}\widetilde{X}^{(N)}_{D,i}\right)^{\ot k}\nonumber\\
&&\left.+\ket{
1}\!\bra{0}^{\ot N}\ot\left(\sum_{i=1}^{N}\widetilde{X}^{(N)\dagger}_{D,i}\right)^{\ot k}\right],
\end{eqnarray}
with the normalization factor given by
\begin{equation}\label{Constr2Norm}
\widetilde{\mathcal{N}}_{D,N}^{(k)}=2\left(ND\sqrt{D}\right)^{k}+2ND^{k}\left[1+(N-1)\sqrt{D}\right]^{k}.
\end{equation}
Notice that as previously mentioned, the LOCC protocol should be
modified in case when $\widetilde{X}^{(N)}_{D}$ follows from in
general unitary $U_{D}$. Due to the modification of the LOCC
protocol, the probability of obtaining
$\widetilde{\Theta}_{\mathsf{AA'}}^{(N,k)}$ in the case of unitary
and unitary Hermitian $U_{D}$ is different. Namely, in the case of
unitary Hermitian matrices amounts to
\begin{equation}\label{prob1}
\widetilde{p}_{D,N}^{(k,1)}=2^{k-1}\widetilde{\mathcal{N}}_{D,N}^{(k)}/
\big(\widetilde{\mathcal{N}}_{D,N}^{(1)}\big)^{k},
\end{equation}
while in the case of unitary non Hermitian the probability of
success is considerably smaller and is given by
\begin{equation}\label{prob2}
\widetilde{p}_{D,N}^{(k,2)}=\widetilde{\mathcal{N}}_{D}^{(N)}/
\big(\widetilde{\mathcal{N}}_{D}^{(N)}\big)^{k}.
\end{equation}

Now the multipartite privacy squeezing (see Section
\ref{PrivacySqueezing}) allows us to change blocks in Eq.
\eqref{SecondConstrLOCC} with their norms, obtaining
\begin{eqnarray}\label{PrivSqueez}
\widetilde{\theta}_{\mathsf{A}}^{(N,k)}&\nmss=\nmss&\frac{1}{\widetilde{\mathcal{N}}^{(k)}_{D,N}}
\left[\sum_{j=0}^{N}\left(\mathcal{P}_{j}^{(N)}+\overline{\mathcal{P}}_{j}^{(N)}\right)
\norsl{\sum_{i=1}^{N}\left|\widetilde{X}^{(N)T_{j}}_{D,i}\right|}^{ k}\right.\nonumber\\
&&\left.+\left(\ket{ 0}\bra{1}^{\ot N}
+\ket{1}\bra{0}^{\ot N}\right)\norsl{\sum_{i=1}^{N}\widetilde{X}^{(N)}_{D,i}}^{k}\right].
\end{eqnarray}

Calculating the respective norms in the above, one may rewrite
Eq. \eqref{PrivSqueez} as
\begin{eqnarray}\label{PrivSq2}
\widetilde{\theta}_{\mathsf{A}}^{(N,k)}&\nmss=\nmss&\frac{D^{k}}{\widetilde{\mathcal{N}}_{D,N}^{(k)}}
\left[ 2(N\sqrt{D})^{k}P_{2,N}^{(+)}\right.\nonumber\\
&&\left.+[1+(N-1)\sqrt{D}]^{k}\sum_{j=1}^{N}
\left(\mathcal{P}_{j}^{(N)}+\overline{\mathcal{P}}_{j}^{(N)}\right)\right].\nonumber\\
\end{eqnarray}
From Eqs. \eqref{Constr2Norm} and \eqref{PrivSq2} one easily
infers that $\widetilde{\theta}_{\mathsf{A}}^{(N,k)}\to
P_{2,N}^{(+)}$ for $k\to\infty$ for any $D\geq 2$, which by virtue
of Theorem III.3 means that the recursive protocol when applied to
copies of $\widetilde{\varrho}_{\mathsf{AA}'}^{(D,N)}$ produces
a state that is arbitrarily close to some multipartite pdit in the
limit of $k\to\infty$. In fact, as the probabilities of success
$\widetilde{p}_{D,N}^{(k,1)}$ and $\widetilde{p}_{D,N}^{(k,2)}$
(see Eqs. \eqref{prob1} and \eqref{prob2}) are positive, according
to the definition of $K_{D}$ (see Definition IV.1) the above
method leads to distillation of secure key from
$\widetilde{\varrho}_{\mathsf{AA}'}^{(D,N)}$. Below we provide
also plots of lower bounds on $K_{D}$ of
$\widetilde{\varrho}_{\mathsf{AA}'}^{(D,N)}$.

\begin{figure}[h!]
\centering{\includegraphics[width=6cm]{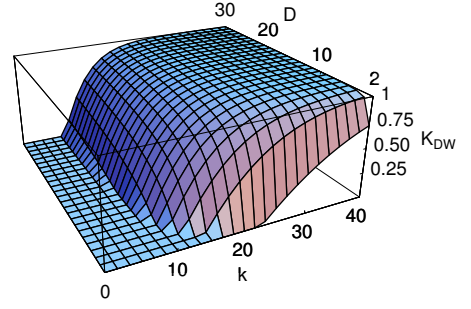}}
\caption{The function appearing on the right--hand side of Eq.
\eqref{Klucz2Constr} (denoted here by $K_{DW}$) in the function of
number of copies $k$ and the dimension $D$. Zero is put whenever
the function is less than zero. Notice that both the parameters
$k$ and $D$ are discrete, however, continuous plot is made to
indicate better the behavior of $K_{DW}$. It is clear from the
plot that for larger $k$ the distillable key of
$\widetilde{\theta}_{\mathsf{A}}^{(N,k)}$ approaches one bit (this
is actually a maximal value obtainable from two--qubit states) and
the convergence depends on $D$. Namely, for higher dimensions $D$
the convergence to the maximal value is
faster.}\label{Constr3Fig1}
\end{figure}

For this purpose we can find the purification of
$\widetilde{\varrho}_{\mathsf{AA}'}^{(D,N)}$ and then the
\textsf{c}q state in the standard basis. The latter has the form
\begin{eqnarray}
\hspace{-0.5cm}\widetilde{\Theta}_{\mathsf{A}E}^{(\mathsf{c}\mathrm{q})}&\nmss=\nmss&a_{D,N}^{(k)}
R_{2}^{(N)}\ot\proj{E_{0}}
+b_{D,N}^{(k)}\nonumber\\
&&\hspace{-0.5cm}\times\sum_{j=1}^{N}\left(\mathcal{P}_{j}^{(N)}\ot\proj{E_{j}}+
\overline{\mathcal{P}}_{j}^{(N)}\ot\proj{\overline{E}_{j}}\right),
\end{eqnarray}
where $\ket{E_{0}}$, $\ket{E_{j}}$, and $\ket{\overline{E}_{j}}$
$(j=1,\ldots,N)$ are orthonormal states kept by Eve, and
coefficients $a_{D,N}^{(k)}$ and $b_{D,N}^{(k)}$ are given by
\begin{equation}
a_{D,N}^{(k)}=\frac{(ND\sqrt{D})^{k}}{\widetilde{\mathcal{N}}_{D,N}^{(k)}}
\end{equation}
and
\begin{equation}
b_{D,N}^{(k)}=\frac{D^{k}}{\widetilde{\mathcal{N}}_{D,N}^{(k)}}[1+(N-1)\sqrt{D}]^{k}.
\end{equation}
One can see from the above that the limit of $k\to \infty$ leads
us to the ideal \textsf{c}q state. Now we can apply the bound
given in Eq. \eqref{DWbound3}. It is easy to verify that all
the quantities $I(A_{i}\!:\!A_{j})$ are equal here (the same holds
for $I(A_{i}\!:\!E)$) and therefore we can rewrite Eq.
\eqref{DWbound3} as
\begin{equation}\label{Klucz2Constr}
K_{D}(\widetilde{\Theta}_{\mathsf{AA}'}^{(N,k)})\geq I(A_{1}\!:\!A_{2})
(\widetilde{\Theta}_{A_{1}A_{2}E}^{(\mathrm{ccq})})-I(A_{1}:E)(\widetilde{\Theta}_{A_{1}A_{2}E}^{(\mathrm{ccq})})
\end{equation}
Exemplary plot of the function appearing on the right--hand side
of Eq. \eqref{Klucz2Constr} (denoted as $K_{DW}$) is presented in
Figure \ref{Constr3Fig1}.
%
%
The behavior of $K_{DW}$ (see Fig. \ref{Constr3Fig1}) confirms the
previous analysis, namely, the more copies we spend the closer
the state is to some multipartite private state we obtain using the
recursive protocol. Thus the higher key rate we can get from the
obtained state $\widetilde{\Theta}_{\mathsf{AA}'}^{(N,k)}$.

We can also get a lower bound on distillable key of the initial
states $\widetilde{\varrho}_{\mathsf{AA}'}^{(D,N)}$. Here we need
to take into account the probability of success
($\widetilde{p}_{D,N}^{(k,1)}$ and $\widetilde{p}_{D,N}^{(k,2)}$)
in the recursive protocol.

\begin{figure}[h!]
(a)\includegraphics[width=6cm]{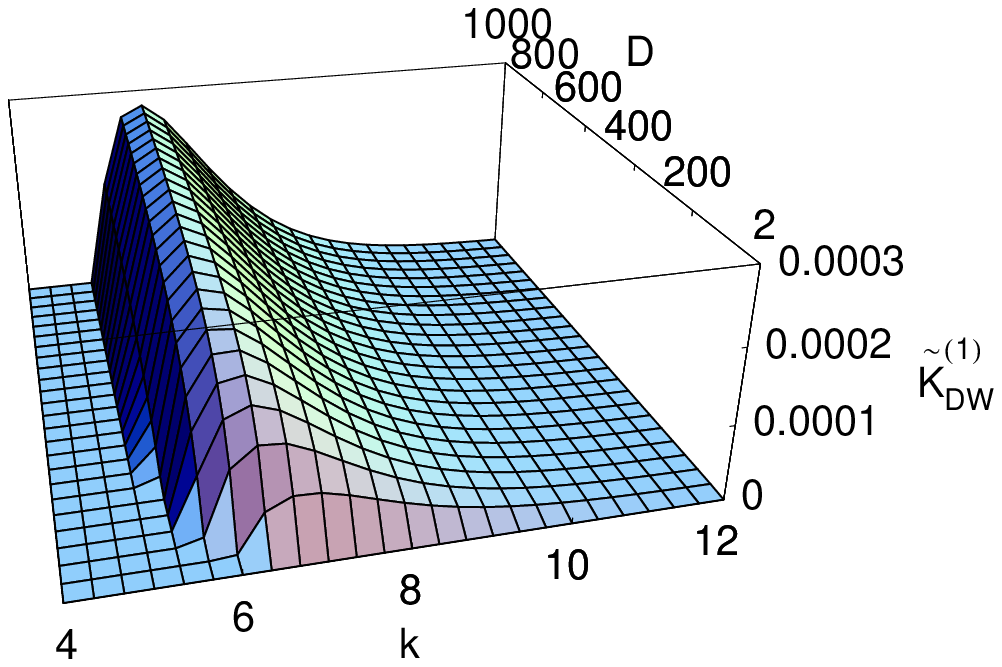}\\
(b)\includegraphics[width=6cm]{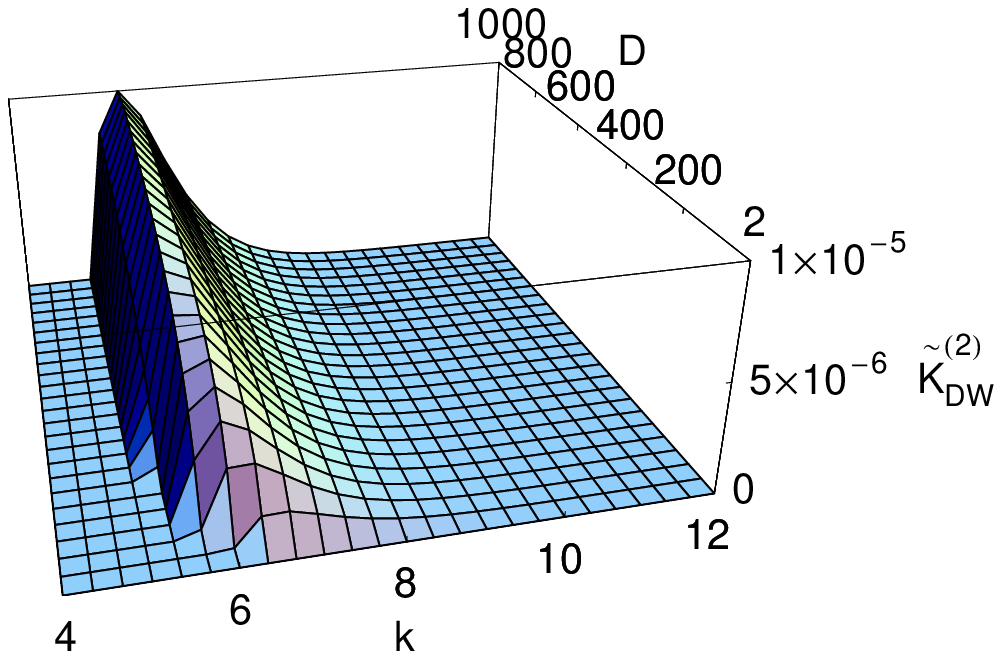} \caption{Lower bounds
on $K_{D}$ of $\widetilde{\varrho}_{\mathsf{AA}'}^{(D,N)}$ in the
function of $k$ and $D$. The upper plot (a) presents lower bound
(denoted here by $\widetilde{K}_{DW}^{(1)}$) on $K_{D}$ in the
case of unitary Hermitian matrices $U_{D}$, while in the second
plot (b) lower bound ($\widetilde{K}_{DW}^{(2)}$) in the case of
unitary but not Hermitian matrices is given. Both are just a
product of probability $\widetilde{p}_{D,N}^{(k1)}$ (left) or
$\widetilde{p}_{D,N}^{(k,2)}$ (right) and $K_{DW}$ plotted in
Figure \ref{Constr3Fig1}. One infers that in the case of unitary
but not Hermitian matrix $U_{D}$ the region of nonzero values of
the plotted function is wider than in the case of unitary
Hermitian matrices.} \label{Constr3Fig2}
\end{figure}

The corresponding bounds
on the distillable keys of
$\widetilde{\varrho}_{\mathsf{AA}'}^{(D,N)}$ are
\begin{eqnarray}
K_{D}^{(1(2))}(\widetilde{\varrho}_{\mathsf{AA}'}^{(D,N)})&\nmss\geq\nmss & \widetilde{p}^{(k,1(2))}_{D,N}\left[I(A_{1}\!:\!A_{2})(\widetilde{\Theta}_{A_{1}A_{2}E}^{(\mathrm{ccq})})\right.\nonumber\\
&&\left.-I(A_{1}\!:\!E)(\widetilde{\Theta}_{A_{1}A_{2}E}^{(\mathrm{ccq})})\right].
\end{eqnarray}
Exemplary plots of the right--hand side of the above (denoted by
$\widetilde{K}_{DW}^{(1(2))}$) both in the case of a unitary
Hermitian matrix (e.g. $\widetilde{V}_{2}^{\ot k}$) and only a
unitary matrix (e.g. $\widetilde{V}_{D}$) are given in Figure
\ref{Constr3Fig2}a and \ref{Constr3Fig2}b.
%
%

%
\section{Remarks on limitations in multipartite quantum cryptography}

So far, we discussed the general scheme allowing for distilling
secure key from multipartite states. It is desirable however to
discuss also what are the limitations of multipartite secure key
distillation.

An interesting effect, which we shall recall here, was reported in
Ref. \cite{RAPH1}, namely, it was shown that though maximal
violation of some Bell inequality it is impossible to distill
secure key from the so--called Smolin state \cite{Smolin}:
\begin{equation}
\varrho^{S}=\frac{1}{4}\sum_{i=0}^{3}\proj{\psi_{i}^{B}}\ot\proj{\psi_{i}^{B}},
\end{equation}
where $\ket{\psi_{i}^{B}}$ $(i=0,\ldots,3)$ are the so--called Bell
states given by
%
$\ket{\psi_{0(1)}^{B}}
=(1/\sqrt{2})\,\left(\ket{01}\pm\ket{10}\right)$
and $\ket{\psi_{2(3)}^{B}}
=(1/\sqrt{2})\,\left(\ket{00}\pm\ket{11}\right)$.
%
This conclusion may be also inferred for the generalizations of
the Smolin state provided in Ref. \cite{RAPH2} and independently
in Ref. \cite{GSS2} (see also Ref. \cite{WangYing} for further
generalizations). These are states of the form
\begin{eqnarray}\label{construction}
&&\varrho_{2}=\proj{\psi_{0}^{B}},\nonumber\\
&&\varrho_{4}^{S}=\frac{1}{4}\sum_{m=0}^{3}U_{2}^{(m)}\varrho_{2}U_{2}^{(m)\dagger}\ot
U_{2}^{(m)}\varrho_{2}U_{2}^{(m)\dagger}\equiv \varrho^{S},\nonumber\\
&&\varrho_{6}^{S}=\frac{1}{4}\sum_{m=0}^{3}U_{4}^{(m)}\varrho_{4}U_{4}^{(m)\dagger}\ot
U_{2}^{(m)}\varrho_{2}U_{2}^{(m)\dagger},\nonumber\\
&&\vdots\nonumber\\
&&\varrho_{2(n+1)}^{S}=\frac{1}{4}\sum_{m=0}^{3}U_{2n}^{(m)}\varrho_{2n}U_{2n}^{(m)\dagger}\ot
U_{2}^{(m)}\varrho_{2}U_{2}^{(m)\dagger}\nonumber\\
\end{eqnarray}
with $U^{(m)}_{k}=\mathbbm{1}_{2}^{\ot (k-1)}\ot \sigma_{m}$
($m=0,\ldots,3$ and $k=2,\ldots$), where $\sigma_{m}$ $(m=1,2,3)$
denote the usual Pauli matrices and $\sigma_{0}=\mathbbm{1}_{2}$.
The state $\varrho_{2}$ is just one of the Bell states, while
$\varrho_{4}^{S}$ is the Smolin state. All states $\varrho_{2n}$
for $n\geq 2$ are bound entangled and for suitable choice of local
observables all states for $n\geq 1$ violate the Bell inequality
\begin{equation}\label{BellInequality}
|E_{1\ldots 11}+E_{1\ldots 12}+E_{2\ldots 21} -E_{2\ldots22}|\le
2
\end{equation}
maximally ($E$ denotes the so--called correlation function, i.e.,
an average of products of local measurement outcomes taken over
many runs of experiment). On the other hand, due to the results of
Refs. \cite{Curty1,Curty2}, and Ref. \cite{PH_przegladowka}, one
may show that it is impossible to distill multipartite secure key
from states $\varrho_{2n}^{S}$ for $n\geq 2$. This shows that
bipartite Ekert protocol [2] cannot be straightforwardly
generalized to multipartite scenario since as discussed above the
maximal violation of most natural multipartite analog of the
CHSH--Bell inequality [38] does not imply nonzero secret key rate,
whereas maximal violation of CHSH--Bell inequality by two qubits
guarantees secrecy. Still, it would be an interesting problem for
further research to identify all Bell inequalities that do the job
in multipartite case as CHSH--Bell inequality does in the case of
two qubits. It should be stressed that some achievements in a
similar direction were already obtained in Refs.
\cite{ScaraniGisin,Sens}, where it was shown that violation of
some Bell inequalities is sufficient condition for security of
multipartite secret sharing protocols \cite{HilleryBuzek} under an
individual attack of some external party.





\section{Conclusions}
Quantum cryptography beyond entanglement distillation is a very
young subject. Until recent times it was natural to expect that
the latter is impossible. While there were significant
developments concerning the bipartite scenario the general
formulation for multipartite case was missing. The present paper
fills this gap by not only generalizing the scheme, but also by
providing new constructions of multipartite bound entangled states
which is really nontrivial. However, there are many unsolved
questions. First it seems to be true that the unconditional
security proof \cite{UnconditionalSecurity} can be extended here
at a cost of the number of estimated local observables, but an
exact analysis of this issue is needed. Moreover, given a fixed
number of parties, it is not known what is the minimal dimension
of elementary system of PPT like bound entangled state that allows
one-way secure key distillation. Does it increase with number of
particles and if so - how the dependence looks like? Are there
bound entangled states with multipartite cryptographic key with
underlying structure corresponding to other classes of pure states
like graph states (see Ref. \cite{Graph})? One may ask why we have
considered only bound entanglement in multipartite scenario. This
is when it is necessary to apply the generalized scheme. Otherwise
qualitatively (though may be not quantitatively -- see subsequent
discussion) just pure entanglement distillation is a sufficient
tool. Quite natural is a question of interplay between the two
approaches in distilling key -- to what extent can we abandon
distillation of p--dits? Finally, can the two processes always be
separated in optimal key distillation scheme: in a sense that one
gets some number of singlet states and some large p--dit which is
bound entangled)? If it were so, the two parts might serve as a
natural measures of free and bound entanglement in the system.
Most likely this is impossible, but one needs a proof. The closely
related question is the one concerning lockability of the secure
key $K_{D}$ (note that nonadditivity of $K_{D}$ has been proved
very recently in Ref. \cite{LiWinter}). While this seems to be a
very hard question in case of bipartite states (though lockability
with respect to Eve has already been ruled out in Ref.
\cite{Christandl}) it may happen to be easier within the
multipartite paradigm presented here (in analogy to classical
bound information which is known only in asymptotic bipartite form
\cite{BoundInfoBipartite} but naturally emerges form bound
entanglement in multipartite case \cite{BoundInfoMultipartite}. In
this context novel upper and lower bounds on $K_D$ are needed (for
recent development see Ref. \cite{MultiSquashed}). This point is
also interesting from the point of view of entanglement as $K_{D}$
is also an entanglement measure.
Further analysis of $K_{D}$ and finding its multi--coordinate
extensions to help in characterization of multipartite
entanglement seems to be rich program for future research.

Also, though in the present paper we are concerned with a
general problem of two--way distillabillity of secure key, it is
interesting to discuss the problem in the context of one--way
schemes. For instance one could ask about bounds on key within
such schemes (see e.g. Ref. \cite{MCL}). On the other hand, it
would be desirable to discuss the present approach in the context
of secure key distillation from continuous variables systems (see
e.g. Refs. \cite{NavascuesBae,2xN,Garcia-Patron}). For instance, it was shown in
Ref. \cite{Navascues} that the generalized version of the protocol
from Ref. \cite{NavascuesBae} does not allow for secure key
distillation from bound entangled states. It seems that within the
multipartite scenario the problem could be simpler a little as one
can have bound entangled multipartite states with some of its
partitions being still NPT.

Needless to say due to Choi--Jamio\l{}kowski isomorphism
\cite{ChJ} the present analysis provides new classes of multiparty
quantum channels for which natural questions on superactivation of
the type found in bipartite case \cite{SmithYard} and other
possible effects of similar type arise.

\acknowledgements This work was supported by UE IP project SCALA.
Partial support from LFPPI network is also acknowledged. R.A.
gratefully acknowledges the support from Ingenio 2010
QOIT and Foundation for Polish Science.


\appendix

\section{Some useful lemmas}
{\it Lemma A.1.} Assume that a given $d\times d$ matrix $B$ is normal.
If $A\geq |B|$ then the matrices
\begin{equation}
\mathcal{M}_{N}(A,B)=\left[
\begin{array}{cccc}
(N-1)A & B  &\ldots & B \\
B^{\dagger} & (N-1)A & \ldots & B\\

\vdots      & \vdots    & \ddots & \vdots\\
B^{\dagger} & B^{\dagger}& \dots &    (N-1)A
\end{array}
\right]
\end{equation}
and
\begin{equation}
\widetilde{\mathcal{M}}_{N}(A,B)=\left[
\begin{array}{ccccc}
(N-1)A      & B & B &  \ldots & B \\
B^{\dagger} & A & 0 &  \ldots & 0\\
B^{\dagger} & 0 & A & \ldots  & 0\\

\vdots & \vdots & \vdots & \ddots & \vdots\\
B^{\dagger}  & 0 & 0 & \ldots & A
\end{array}
\right].
\end{equation}
are positive.

{\it Proof.} We prove the lemma only for $\mathcal{M}_{N}(A,B)$ as
the proof for $\widetilde{\mathcal{M}}_{N}(A,B)$ goes along the
same lines.

The matrix $\mathcal{M}_{N}(A,B)$
consists of $N^{2}$ blocks $d\times d$ each and consequently the whole
matrix has the dimensions $Nd\times Nd$. Thus
to prove positiveness we need to show that for any
$\ket{\Psi}\in\mathbb{C}^{Nd}$ the inequality $\bra{\Psi}\mathcal{M}_{N}(A,B)\ket{\Psi}\geq 0$ holds.
It is clear that an arbitrary vector $\ket{\Psi}\in\mathbb{C}^{Nd}$ may be written as
\begin{equation}\label{WconstrLemat3}
\ket{\Psi}=\left[
\begin{array}{c}
\ket{x_{1}}\\
\vdots\\
\ket{x_{N}}
\end{array}
\right],
\end{equation}
where each $\ket{x_{i}}$ belongs to $\mathbb{C}^{d}$. Then a rather straightforward algebra
leads to
\begin{eqnarray}
\bra{\Psi}\mathcal{M}_{N}(A,B)\ket{\Psi}&\nmsss=\nmsss&(N-1)\sum_{i=1}^{N}\bra{x_{i}}A\ket{x_{i}}\nonumber\\
&&+2\sum_{\substack{i,j=1\\i<j}}^{N}\mathrm{Re}\bra{x_{i}}B\ket{x_{j}}.
\end{eqnarray}
By virtue of the assumption that $A\geq |B|$ and the inequality $\mathrm{Re}z\geq -|z|$
satisfied for any $z\in\mathbb{C}$, one has
\begin{eqnarray}\label{WconstrLemat1}
\bra{\Psi}\mathcal{M}_{N}(A,B)\ket{\Psi}&\nmsss\geq\nmsss&
(N-1)\sum_{i=1}^{N}\bra{x_{i}}|B|\ket{x_{i}}\nonumber\\
&&-2\sum_{\substack{i,j=1\\i<j}}^{N}\left|\bra{x_{i}}B\ket{x_{j}}\right|,
\end{eqnarray}
Since $B$ is assumed to be a normal matrix it may be given as $B=\sum_{k}\lambda_{k}\proj{\varphi_{k}}$
with $\{\lambda_{k}\}$ being, in general, the complex eigenvalues of $B$, while $\{\ket{\varphi_{k}}\}$
its orthonormal eigenvectors. Putting the spectral decomposition of $B$ into
Eq. \eqref{WconstrLemat1}, introducing
$a_{ik}=|\langle x_{i}|\varphi_{k}\rangle|\geq 0$, and taking into account
the fact that $|\sum_{i}\xi_{i}|\leq \sum_{i}|\xi_{i}|$, we obtain
\begin{equation}\label{WconstrLemat2}
\bra{\Psi}\mathcal{M}_{N}(A,B)\ket{\Psi}\geq \sum_{k}|\lambda_{k}|
\left(\sum_{i=1}^{N}a_{ik}^{2}-2\sum_{\substack{i,j=1\\i<j}}^{N}a_{ik}a_{jk}\right).
\end{equation}
It is clear from Eq. \eqref{WconstrLemat2} that to show nonnegativity of
$\bra{\Psi}\mathcal{M}_{N}(A,B)\ket{\Psi}$ for any $\ket{\Psi}\in\mathbb{C}^{Nd}$,
one has to prove that for all $k$ the term in brackets in Eq. \eqref{WconstrLemat2}
is nonnegative. This, however, follows from the fact that
\begin{equation}
\sum_{\substack{i,j=1\\i<j}}^{N}(a_{ik}-a_{jk})^{2}\geq 0,
\end{equation}
finishing the proof. $\blacksquare$\\

{\it Lemma A.2.} Let $A=\sum_{i,j=0}^{d-1}a_{i}^{j}\ket{i}\bra{j}$
be a positive matrix obeying $\Tr A\leq 1$. Assume that each
element of $A$ lying in $i$th row (and $i$th column due to
hermiticity of $A$) is close to $1/d$ in the sense that it obeys
$|a_{i}^{j}-1/d|\leq \epsilon$ for some $1/d>\epsilon>0$. Then
$|a_{i}^{j}-1/d|\leq\eta(\epsilon)$ for any $i,j=0,\ldots,d-1$,
where $\eta(\epsilon)\to 0$ for $\epsilon\to 0$.

{\it Proof.} The proof is rather technical and we present only its
sketch here (detailed proof may be found in Ref.
\cite{DoktoratRA}). First, let us fix the chosen row to be the
first one, i.e., $i=0$. Then, from the positivity of $A$ it
follows that any matrix of the form
\begin{equation}
\left[
\begin{array}{cc}
a_{0}^{0} & a_{0}^{j}\\[1ex]
a_{0}^{j*} & a_{j}^{j}
\end{array}
\right]
\end{equation}
is positive. Now, from its positivity we have that
$a_{0}^{0}a_{j}^{j}\geq \left|a_{0}^{j}\right|^{2}$, which
together with the assumption that $a_{0}^{0}$ and $a_{0}^{j}$ are
close to $1/d$ and $\epsilon<1/d$, implies that $a_{j}^{j}$ must
obey $a_{j}^{j}\geq 1/d-3\epsilon$ for any $j=1,\ldots,d-1$.
Taking into account the assumption that $\Tr A\leq 1$, one also
has that each $a_{j}^{j}$ must be bounded from above as
%
$a_{j}^{j}\leq 1/d+3(d-1)\epsilon$ for $j=1,\ldots,d-1$.
%
Therefore, we have that all diagonal elements of $A$ satisfy
\begin{equation}\label{Appendix1}
\left|a_{j}^{j}-\frac{1}{d}\right|\leq \alpha(\epsilon)
\end{equation}
with $\alpha(\epsilon)\to 0$ for $\epsilon\to 0$. Now, we need to
prove that the remaining off--diagonal elements of $A$ are also
close to $1/d$. For this purpose let us notice that from the fact
that $A\geq 0$ one has that the following matrices
\begin{equation}
\left[
\begin{array}{ccc}
a_{0}^{0} & a_{0}^{i} & a_{0}^{j} \\[1ex]
a_{0}^{i*} & a_{i}^{i} & a_{i}^{j}\\[1ex]
a_{0}^{j*} & a_{i}^{j*} & a_{j}^{j}
\end{array}
\right] \qquad (0<i<j)
\end{equation}
are also positive. Since we can now say that all elements in the
first row (and column) and all the diagonal elements obey
(\ref{Appendix1}), it follows, after some technical calculations,
that $a_{i}^{j}$ has to satisfy such inequality, however, with
some other function which vanishes for $\epsilon\to 0$. Finally,
we have that any of the elements of $A$ satisfies
\begin{equation}
\left|a_{i}^{j}-\frac{1}{d}\right|\leq \eta(\epsilon) \qquad
(i,j=0,\ldots,d-1)
\end{equation}
with $\eta(\epsilon)\to 0$ whenever $\epsilon\to 0$.

Of course, we can always assume that the elements in some fixed
row of $A$ is bounded by different $\epsilon$s. Then, however, we
can take the largest one. $\blacksquare$

\end{document}